\documentclass[epjH]{svjour}
\usepackage{etoolbox}
\newcommand*{\affaddr}[1]{#1} 
\newcommand*{\affmark}[1][*]{\textsuperscript{#1}}
\usepackage{graphicx}

\usepackage{soul}

\usepackage{silence}
\usepackage[style=base, labelsep=space, labelfont=bf]{caption}
\usepackage{amsmath}
\usepackage{amssymb}
\usepackage[latin1]{inputenc}
\usepackage [noconfigs, USenglish]{babel}
\usepackage [autostyle, english = american]{csquotes}
\MakeOuterQuote{"}

\usepackage{footnote}

\usepackage[]{threeparttable}

\usepackage{isotope}

\usepackage{pdflscape}

\journalname{European Physical Journal}

\usepackage[
    natbib=true,
    maxcitenames=2,
    style=authoryear, 
    bibstyle=authoryear,
    backend=biber, 
    ]{biblatex}
\bibliography{Borradores}

\DeclareFieldFormat[article]{title}{\mkbibquote{#1}}
\renewbibmacro*{in:}{%
  \ifentrytype{article}
    {}
    {\printtext{\bibstring{in}\intitlepunct}}}

\newcommand{\comentario}[1]{}

\usepackage{upgreek}

\usepackage{url}

\usepackage{tikz-cd}
\tikzcdset{arrow style=tikz}

\PassOptionsToPackage{bookmarks=true,bookmarksdepth=2}{hyperref}

\usepackage{soul}
\usepackage{pdfcomment} 
\pdfcommentsetup{final} 
\pdfcommentsetup{color={1.0 1.0 0.0}, open=true}

\hypersetup{colorlinks=true, pdfstartview=FitH} 

\DeclareCiteCommand{\citeyearpar}[\mkbibparens]
    {\usebibmacro{prenote}}
    {\bibhyperref{\printfield{year}}\bibhyperref{\printfield{extradate}}}
    {\multicitedelim}
    {\usebibmacro{postnote}}

\begin{document}
\title{Blurred orbits and blurred particles: Heisenberg's 1926 helium atom
}
\author{%
Gonzalo Gimeno\protect\affmark[1] \and  Mercedes Xipell\affmark[2] \and Enric Pérez\affmark[3]
}
\authorrunning{%
Gonzalo Gimeno \and Mercedes Xipell
}
\institute{
              Gonzalo Gimeno (corresponding author) \href{https://orcid.org/0000-0003-4348-6437}{ORCID Id. 0000-0003-4348-6437} \\
              gonzalo.gimeno@uab.cat
         \and
              Mercedes Xipell \href{https://orcid.org/0000-0001-6930-8642}{ORCID. Id. 0000-0001-6930-8642} \\
              mercedes.xipell@uab.cat
          \and
              Enric Pérez \href{https://orcid.org/0000-0003-0303-4401}{ORCID. Id. 0000-0003-0303-4401} \\
              enperez@ub.edu\\
              \\
          \affaddr{\affmark[1,2] Institut d'Història de la Ciència, Research Module C, Carrer de Can Magrans s/n, Universitat Autònoma de Barcelona, 08193, Bellaterra, Barcelona, Spain.}\\
          \affaddr{\affmark[3]}Departament de Física de la Matèria Condensada, Universitat de Barcelona, Carrer de Martí i Franquès 1, 08028 Barcelona, Spain.}

\abstract{
This work analyses the extent to which the "blurred orbits" of the current model for the atom, drafted by Heisenberg in 1926, fits the image of a bunch of wandering electrons around a nucleus. We will deal with early appearances of the concept of indistinguishable particles within the frame of quantum mechanics. There are few studies on the use of this concept in Heisenberg's 1926 papers on helium, in contrast to the large number of them on its use in Bose--Einstein's 1924 papers. We will discuss to what extent Heisenberg's approach leads to a purely statistical interpretation of this concept. We will also study the viewpoint of Dirac, who dealt with the same topic few months later. Although the indistinguishability of the electrons and indeterminacy are common explanations for the blurring of electron orbits, we argue that such an image is an oversimplification which masks interesting aspects of: a) the dynamics of bound electrons and b) the deeper implications of indistinguishability in quantum mechanics.}
\maketitle

\hyphenpenalty=5000
\tolerance=1000

\setlength{\footskip}{30pt}


\section{Introduction}\label{introduction}

The structure of the atom remains under scrutiny, still keeping most of its mystery even today. Despite advances in measuring physical quantities, there are unanswered questions regarding many atomic components. Muonic versions of light atoms are currently being used to help determine the characteristics of the nucleus. Specifically, a recent result on the radius of the nucleus of He can be seen at \citet{RefWorks:1350}. But the study of such low-stability helium-4 (\isotope[4]{He}) nucleus-muon compounds do not completely explain the stability and properties of their more stable nucleus-electron partners. Significantly, such studies do not refrain from using the classical image of "a single muon [that] orbits a bare nucleus" [ib. p. 528] together, though, with the corresponding muon wave function. This reliance on classical terms (orbits, angular momenta etc.) contrasts with the now usual presentation of atomic electron clouds from the perspective of their quantum properties, the indeterminacy of electron position and the rejection of any pictorial model with electron paths. The importance of chemically stable compounds in chemistry and the need to understand nature starting from classical mechanics justify the use of semiclassical approaches \citetext{\citealt{RefWorks:1433}; \citealt[Sect. IV]{RefWorks:1416}; \citealt{RefWorks:1438}; \citealt{RefWorks:1440}; \citealt{RefWorks:1434}} to overcome these difficulties.

The introduction of Bohr's atomic model in 1913 was successful in many ways but attempting to understand atomic structure via the spatiotemporal description of the nucleus and the orbiting electrons worked well only in the simplest case, namely hydrogen. Similar attempts for the next most simple element, helium, were unsuccessful under Bohr's hypotheses. In the twenties, Landé, Pauli, Born, Heisenberg, and others tried to make sense of Bohr's theory in the case of helium just to conclude the impossibility of a reasonable model that explained the stability of the helium atom, its ionization energy and its spectroscopic terms---notably the double spectra of para- and ortho-helium and the Zeeman multiplets---under orbital assumptions.

It is generally accepted that, among the reasons that forced physicists to abandon the pursuit of a general description of the atom under the Bohr planetary model, were the difficulties of accounting for the known experimental constants of helium. A number of studies examine how Bohr's 1913 model \citetext{\citealt[]{RefWorks:1011}} suffered gradual distrust shortly after it was presented \citetext{\citealt[Ch. 8]{RefWorks:1445}; \citealt[Ch. 7]{RefWorks:1468}}. The distrust grew with the appearance of more sophisticated models, such as the elaboration by Sommerfeld on a previous model by Voigt, Heisenberg's core model, or Landé's vector model. But, despite some of them succeeding in explaining experimental results, they were all found to undermine principles generally accepted at the time.

The old theory failed to justify the spectrum of helium, and to calculate its ground-state energy. None of this could be solved under the assumptions of the Bohr atom \citetext{\citealt[]{RefWorks:1416}}. These contradictions led to an explicit rejection of models (and, as we will argue, to a new approach to indistinguishable particles) that contributed to the birth of the new quantum mechanics.\footnote{Indistinguishability of identical particles was already used in statistical mechanics in the sense of \citet{RefWorks:583}; we hold that a somewhat different concept was being proposed along with the new quantum mechanics. While the term "indistinguishability" is used throughout this work, it has not always had a single, agreed-upon meaning. See \citetext{\citealt{RefWorks:RefID:24-dieks1990quantum}; \citealt{RefWorks:RefID:21-monaldi2009note}; \citealt{RefWorks:RefID:25-versteegh2011gibbs}; \citealt{RefWorks:RefID:1408-dieks2014logic}; \citealt{RefWorks:RefID:23-dieks2021identical}; \citealt{RefWorks:1458}}. Our use of the term throughout this work is mainly in the sense of permutation-invariant particles as is more formally introduced in section \ref{Indist} at p. \pageref{Indist}.} Historians usually place this birth in the publication of Heisenberg's \emph{Umdeutung} paper \citet{RefWorks:58}.

Several authors describe the models \emph{previous} to the birth of the new quantum mechanics and discuss their success and contradictions \citetext{\citealt{RefWorks:1390}; \citealt{RefWorks:1391}; \citealt[p. 203]{RefWorks:1405}; \citealt[Ch. 8.2]{RefWorks:1445}; \citealt[Ch. 7]{RefWorks:1468}}. In this work we analyse how the new quantum mechanics shaped an entirely different atomic model.

The new quantum mechanics, although successful in accounting for some of the unsolved problems of helium, was plagued by side-effects. One of them was the reinforcement of a (then recent) suspicion that particles (including massive particles like electrons) were subject to unexpected correlations.

As we will remark, these correlations had been noticed years ago for the case of light quanta, although the fact that light quanta preserved their wave nature, despite their quantization, made it possible to identify a certain consistency, in the end, a wave can easily overlap with another wave. However, in 1924, Einstein's application of Bose's ideas on light quanta to gas molecules led to an unforeseen property of mass particles: they could be said to be correlated in the same way that light quanta were. When kept within the frame of combinatorics, as was the case of Bose's and Einstein's calculations, the correlations (i.e. the lack of statistical independence) could be understood as implied by the indistinguishability of the particles.

This indistinguishability was not immediately recognized although it is now a frequent (and controversial) interpretation of those correlations. Indistinguishability in its modern form challenges the concept of individual particle as if particles, understood in the classical meaning of the word, do not exist at all, and suggests quantum field theory as the only frame in which particles have some meaning \citetext{\citealt{RefWorks:RefID:24-dieks1990quantum}}. But the need to account for the natural appearance of particles in the experimental practice of physics has been claimed to require some alternative explanation \citetext{\citealt{RefWorks:RefID:23-dieks2021identical}}. The gap between classical physics and quantum physics keeps calling for some bridge.

As claimed by \citet[p. 1148]{RefWorks:1417}, an understanding of the relation between classical and quantum mechanics would be beneficial. Consequently with this claim, a sustained examination how classical mechanics came to be abandoned along with the establishment of quantum mechanics is, therefore, in order. It is our meaning that there are some contributions to this issue that still deserve further analysis. This is the case of what we call Heisenberg's "resonance papers". Under this name, we consider four papers written by Heisenberg in which he deals, from the point of view of the new quantum mechanics, with several problems of the multi-electron atom under the following titles:
\begin{itemize}
  \item "Mehrkörperproblem und Resonanz in der Quantenmechanik" \citetext{\citealt[]{RefWorks:1264}},
  \item "Über die Spektra von Atomsystemen mit zwei Elektronen" \citetext{\citealt[]{RefWorks:1278}},
  \item "Schwankungserscheinungen und Quantenmechanik" \citetext{\citealt[]{RefWorks:1277}}\footnote{Beller calls this the "fluctuation paper" in \citet{RefWorks:38}.}, and
  \item "Mehrkörperprobleme und Resonanz in der Quantenmechanik. II" \citetext{\citealt[]{RefWorks:1276}}.
\end{itemize}
It is delightful to read Kuhn's interviews with Heisenberg in 1963, when he asked about this whole set of papers:
\begin{quote}
  But the first \citetext{\citealt[]{RefWorks:1264}} of the two resonance papers, which is also the first of that whole group, is the one I'd particularly like to talk about because it really contains almost everything that comes later. \emph{I sit there reading it with my mouth hanging open because of the number of new ideas that all of a sudden get tied together here}. \citetext{AHQP, Interview with Heisenberg, February 25, 1963; our emphasis}
\end{quote}

In 1926 and 1927, Heisenberg made some calculations on atoms and molecules using the new formulations of quantum mechanics, then recently suggested by several physicists including himself, Max Born, Pascual Jordan, Paul Dirac, Wolfgang Pauli, and Erwin Schrödinger. Part of his work was gathered in the four papers mentioned; in these, resonance is a key physical assumption. The orthodox view that Heisenberg gave up looking for a spacetime atomic dynamics seems quite certain, but one could come to the easy but wrong conclusion that he discarded the idea of even the existence of such an atomic dynamics.

The resonance papers met with uneven reception. For instance, the first paper \citetext{\citealt[]{RefWorks:1264}} did not get too much attention from physicists and is generally ignored by historians. The main lines of the work and the context in which the paper was produced can be found in \citet{RefWorks:40} and Kuhn's interview with Heisenberg in the AHQP. A more detailed analysis of the work with an explanation of the way Heisenberg justified the spectra of Para- and Ortho-helium has been elaborated at \citet[]{RefWorks:RefID:27-gearhart2010astonishing}, \citet{RefWorks:984} and \citet[]{RefWorks:RefID:19-duncan2023constructing}. While \citet[]{RefWorks:RefID:27-gearhart2010astonishing} initially emphasizes the relevance of indistinguishable particles in Heisenberg's work, he does not develop how Heisenberg used them. \citet[]{RefWorks:RefID:19-duncan2023constructing} give a more detailed account including some results of Heisenberg's paper on two-electron atoms, but leave unanswered how is it that Heisenberg arrived to Bose--Einstein statistics in connection with the problem of bound electrons. This explanation has been only attempted at \citet[]{RefWorks:984}. \citet[]{RefWorks:1458} and \citet[]{RefWorks:1437} briefly analyze this first paper in the context of the origins of quantum statistics.

The paper on the helium atom, \citet[]{RefWorks:1278}, and its extension to lithium and molecules, \citet[]{RefWorks:1276}, have been generally bypassed by historians (although there is an interesting introduction to paper \citet[]{RefWorks:1278} on helium in \citet[]{RefWorks:1281}) but they were well received by physicists and their main results form an essential part of today's atomic theory.

The paper on fluctuations, \citet[]{RefWorks:1277}, seems to have captured the attention of only some historians. Mara Beller, for instance, elaborates repeatedly on the hypothesis that this paper and its "Copenhagen responses" manifested the differing philosophical positions of Schrödinger and the "Copenhagen group" and on how these differences shaped the "quantum revolution" \citetext{\citealt[p. 194]{RefWorks:38}}. But although it deserved Schrödinger's direct response in \citet{RefWorks:1404}, and stimulated Dirac's and Jordan's probabilistic formalism \citetext{\citealt[p. 194]{RefWorks:38}}, it seems to be suppressed by a curtain of silence. Beller focuses mainly on the clash between Heisenberg and Schrödinger about the continuity or discontinuity of physical phenomena \citetext{\citealt[p. 47, pp. 76-77]{RefWorks:38}} in the context of a justification of Heisenberg's uncertainty paper and, more generally, of the justification of the Copenhagen Interpretation of quantum mechanics but, does not connect the work with Heisenberg's papers on helium \citetext{\citealt[ch. 10]{RefWorks:38}}. The issue is all but simple with ramifications in semi-classical approaches and requires a thorough discussion involving a specific analysis, so we will not deal with the paper on fluctuations here. It seems to us that the current, mixed status of the electronic orbits is a sequel of the unsolved debate between the architects of quantum mechanics on the need for observability of physical phenomena.\footnote{The need for observability was claimed, for instance, in Dirac's declaration that the theory should restrict itself to observable quantities: "[the fact that the theory] enables one to calculate just those quantities [frequencies and intensities] that are of physical importance, and gives no information about quantities such as orbital frequencies [is a] \emph{very satisfactory characteristic} to persist in all future developments" \citetext{\citealt[p. 666-667, our emphasis.]{RefWorks:RefID:605-dirac1926theory}}}

The present work corroborates some aspects of the studies previously mentioned, but differs from, and completes them, in that it a) reveals that Heisenberg's contribution to the statistical approach to ideal gases is present in his formulation of the symmetrization requirements for the state function and its connection with the Pauli exclusion principle, but he did not fully establish the link he proposed between the Bose--Einstein counting method and quantum mechanics; b) shows that this link was due to Dirac; c) clarifies the different ways in which indistinguishability could be understood in Heisenberg's papers; d) highlights the connection between Heisenberg's works and the (still undeveloped) concept of quantum superposition. Consequently, our emphasis will be on how indistinguishability and superposition combined to transform a planetary system of material bodies into a blurred mess of wave-like, non-identifiable, indistinguishable, but statistically predictable particles.

This study will focus mainly in a review of the papers "Mehrkörperproblem und Resonanz in der Quantenmechanik" \citetext{\citealt{RefWorks:1264}} and "Über die Spektra von Atomsystemen mit zwei Elektronen", \citetext{\citealt{RefWorks:1278}}. A reasonable comprehension of their relevance needs to begin with some comment on previous and later work.

As previous work we will consider papers by Bose, Einstein and Fermi \citetext{\citealt{RefWorks:1067, RefWorks:1058, RefWorks:1066, RefWorks:1060, RefWorks:RefID:10-fermi1926sulla, RefWorks:RefID:9-fermi1926zur}} on the theory of ideal gases as the frame which suggested to Heisenberg a connection of his matrix mechanics with the Bose--Einstein way of counting. We will show that Heisenberg's claimed connection is justified but inadequate.

We will briefly discuss the similarities and differences between Heisenberg's 1926 papers and a previous work by Born and Heisenberg on the helium atom \citetitle{RefWorks:1393} \citetext{\citealt{RefWorks:1393}}.

Additionally, we analyze their use in the scholarly community through the text  \citetitle{RefWorks:1270} of \citet{RefWorks:1270} and the more consolidated analysis made by \citet{RefWorks:1400} in his book \citetitle{RefWorks:1400} which, to the best of our knowledge, is still considered a standard interpretation of Heisenberg's work.

In this work, an attempt is made to justify the following claims:
\begin{enumerate}
  \item Indistinguishability of identical particles, as used in statistical mechanics, led Heisenberg to the symmetrization postulate: state functions should be symmetric or antisymmetric. This postulate then drove him to:
         \begin{itemize}
           \item a mathematical expression of the Pauli exclusion principle in the form of an antisymmetric global state function which, in turn, allowed him to produce a theoretical explanation of the double spectrum of para- and ortho- helium (singlet and triplet configurations).
           \item the development, under the term "resonance", of a construct previously used by Schrödinger and later known as quantum superposition.
         \end{itemize}
  \item Combinatorial probabilities, employed in statistical mechanics to calculate the configuration of thermodynamical equilibrium, were about to be replaced, in Heisenberg's work, by wave functions as statistical devices for atomic calculations.
  \item Heisenberg used the values of the eigenfunctions, not their number, to calculate the statistical weights, thus diverging from thermodynamics. Consequently, he failed in his effort to justify, using the $1/N!$ factor reduction, the (existing) link between quantum mechanics and Bose--Einstein statistics. Hence, our claim gives support to prior opinions that the final derivation of the Bose--Einstein statistical counting method from a quantum mechanical approach was due to Dirac shortly after (e.g., \citetext{\citealt{RefWorks:984}}).
\end{enumerate}

This paper is organized as follows. Section \ref{1923-1925}--\nameref{1923-1925}, is devoted to the different models of the helium atom that existed (and had been discarded) at the time. We also briefly sketch Heisenberg's lines of work on the helium atom. In Section \ref{1926}--\nameref{1926}, we present Heisenberg's both qualitative (sect. \ref{1926a}) and quantitative (sect. \ref{1926b}) arguments. In Section \ref{Synthesis}--\nameref{Synthesis}, we give a quick reminder of Heisenberg's papers main ideas. In Section \ref{qstats}--\nameref{qstats}, we describe the elements of quantum statistics introduced by Bose, Einstein and Fermi. In Section \ref{indalaH}--\nameref{indalaH}, we highlight Heisenberg's polysemous ideas on indistinguishability of particles and emphasize his reliance on the corpuscular nature of matter. In Section \ref{Aftermath}--\nameref{Aftermath}, we sketch the path followed by the theory in the months subsequent to the publication of the papers. In Section \ref{discussion}--\nameref{discussion}, we examine some difficulties in the interpretation of his works, mostly related to blurred orbits, indistinguishability and superposition, and we also conduct a brief historical review.

\section{The preceding years 1923-1925: The banishing of electron orbits}\label{1923-1925}

Heisenberg was a key figure in undermining the concept of electronic orbits, as physicists increasingly recognized the inadequacy of the Bohr model for atoms beyond hydrogen.\footnote{The reader interested in an extended account of the problems leading to the failure of the orbital representation and its connection with the helium atom will find useful the book of \citet[pp. 313-355]{RefWorks:1445}. The comment at section \ref{ac}-\nameref{ac} applies to this section.}  Van Vleck \citeyearpar[p. 419]{RefWorks:van1922dilemma}, for instance, claimed openly in a brief note that "no satisfactory model of normal helium" had been yet devised. As an example, van Vleck cited the wrong ionization potentials obtained from the models of Bohr, Langmuir, Franck and Reiche, and Landé.

Certainly, once the conceptual problems of the Bohr model had been accepted (or temporarily bypassed), physicists at the time struggled with the theoretical determination of the energy of first ionization of helium. Several models had been proposed, among them  (figure \ref{Hemodels}, also \citet{RefWorks:van1922normal, RefWorks:1433}):
\begin{itemize}
  \item[(a)] the model of \citet[]{RefWorks:1011} with two diametrically opposite electrons orbiting on the same circle in the same direction.
  \item[(b)] a semicircular model by \citet{RefWorks:1429} in which the two electrons go over respective arcs of two coplanar circular orbits in opposite directions, reversing when Coulomb repulsion makes them turn back to trace the same arc until they encounter each other again at the opposite extremum.
  \item[(c)] \label{0050}a model by \citet{RefWorks:1436} with two electrons in the same direction on coplanar circular orbits of different radii.
  \item[(d)] a double-circle model by \citet{RefWorks:1429} with the electrons rotating on the same direction on two circular orbits of the same radius over different parallel planes symmetrically located at certain distance from the nucleus.
  \item[(e)] a model proposed by \citet{RefWorks:1443} and supported by \citet{RefWorks:1431} and \citet{RefWorks:1432} with the electrons on two non-coplanar circular orbits intersecting at a certain angle, and synchronized to cross the intersections at the same time over the respective orbits. This model was eventually calculated by  \citet{RefWorks:van1922normal}.
\end{itemize}

\begin{figure}[ht]
\centering
  \caption{Examples of periodic configurations of the electron pair in helium that served as classical models for the ground state: (a) Bohr, 1913; (b) and (d) Langmuir, 1921; (c) Landé, 1919; (e) Kemble, 1921 and Kramers, 1923 (See \citet{RefWorks:1433})}
  \label{Hemodels}       

  \begin{center}
    \tikzset{every picture/.style={line width=0.75pt}} 
    \begin{tikzpicture}[x=0.75pt,y=0.75pt,yscale=-1,xscale=1]

    \draw  [line width=1.5]  (18.77,50.77) .. controls (18.77,29.01) and (36.85,11.38) .. (59.15,11.38) .. controls (81.46,11.38) and (99.53,29.01) .. (99.53,50.77) .. controls (99.53,72.53) and (81.46,90.17) .. (59.15,90.17) .. controls (36.85,90.17) and (18.77,72.53) .. (18.77,50.77) -- cycle ;
    \draw  [dash pattern={on 4.5pt off 4.5pt}]  (25.53,71.27) -- (92.2,28.2) ;
    \draw  [dash pattern={on 4.5pt off 4.5pt}]  (18.77,50.77) -- (99.53,50.77) ;
    \draw  [fill={rgb, 255:red, 0; green, 0; blue, 0 }  ,fill opacity=1 ] (56.6,50.77) .. controls (56.6,49.4) and (57.74,48.28) .. (59.15,48.28) .. controls (60.56,48.28) and (61.71,49.4) .. (61.71,50.77) .. controls (61.71,52.15) and (60.56,53.26) .. (59.15,53.26) .. controls (57.74,53.26) and (56.6,52.15) .. (56.6,50.77) -- cycle ;
    \draw  [fill={rgb, 255:red, 0; green, 0; blue, 0 }  ,fill opacity=1 ] (89.65,28.2) .. controls (89.65,26.83) and (90.79,25.71) .. (92.2,25.71) .. controls (93.61,25.71) and (94.75,26.83) .. (94.75,28.2) .. controls (94.75,29.57) and (93.61,30.69) .. (92.2,30.69) .. controls (90.79,30.69) and (89.65,29.57) .. (89.65,28.2) -- cycle ;
    \draw  [fill={rgb, 255:red, 0; green, 0; blue, 0 }  ,fill opacity=1 ] (22.98,71.27) .. controls (22.98,69.9) and (24.12,68.78) .. (25.53,68.78) .. controls (26.94,68.78) and (28.08,69.9) .. (28.08,71.27) .. controls (28.08,72.65) and (26.94,73.76) .. (25.53,73.76) .. controls (24.12,73.76) and (22.98,72.65) .. (22.98,71.27) -- cycle ;
    \draw  [draw opacity=0][line width=1.5]  (139.82,40.04) .. controls (145.5,25.66) and (159.19,15.52) .. (175.18,15.52) .. controls (191.17,15.52) and (204.86,25.66) .. (210.54,40.04) -- (175.18,54.86) -- cycle ; \draw  [line width=1.5]  (139.82,40.04) .. controls (145.5,25.66) and (159.19,15.52) .. (175.18,15.52) .. controls (191.17,15.52) and (204.86,25.66) .. (210.54,40.04) ;
    \draw  [draw opacity=0][line width=1.5]  (210.66,63.17) .. controls (204.98,77.55) and (191.29,87.69) .. (175.29,87.69) .. controls (159.31,87.69) and (145.61,77.56) .. (139.93,63.18) -- (175.29,48.35) -- cycle ; \draw  [line width=1.5]  (210.66,63.17) .. controls (204.98,77.55) and (191.29,87.69) .. (175.29,87.69) .. controls (159.31,87.69) and (145.61,77.56) .. (139.93,63.18) ;
    \draw  [fill={rgb, 255:red, 0; green, 0; blue, 0 }  ,fill opacity=1 ] (192.58,21.69) .. controls (192.58,20.22) and (193.66,19.02) .. (194.99,19.02) .. controls (196.33,19.02) and (197.41,20.22) .. (197.41,21.69) .. controls (197.41,23.16) and (196.33,24.35) .. (194.99,24.35) .. controls (193.66,24.35) and (192.58,23.16) .. (192.58,21.69) -- cycle ;
    \draw  [fill={rgb, 255:red, 0; green, 0; blue, 0 }  ,fill opacity=1 ] (191.96,82.02) .. controls (191.96,80.55) and (193.04,79.36) .. (194.37,79.36) .. controls (195.71,79.36) and (196.79,80.55) .. (196.79,82.02) .. controls (196.79,83.49) and (195.71,84.68) .. (194.37,84.68) .. controls (193.04,84.68) and (191.96,83.49) .. (191.96,82.02) -- cycle ;
    \draw  [dash pattern={on 4.5pt off 4.5pt}]  (176.03,88.22) -- (176.6,17.44) ;
    \draw  [fill={rgb, 255:red, 0; green, 0; blue, 0 }  ,fill opacity=1 ] (173.9,52.2) .. controls (173.9,50.73) and (174.98,49.54) .. (176.32,49.54) .. controls (177.65,49.54) and (178.73,50.73) .. (178.73,52.2) .. controls (178.73,53.67) and (177.65,54.86) .. (176.32,54.86) .. controls (174.98,54.86) and (173.9,53.67) .. (173.9,52.2) -- cycle ;
    \draw  [dash pattern={on 4.5pt off 4.5pt}]  (176.76,21.19) -- (193.99,21.06) ;
    \draw  [dash pattern={on 4.5pt off 4.5pt}]  (177.14,82.78) -- (194.37,82.65) ;
    \draw  [line width=1.5]  (19.34,153.56) .. controls (19.34,133.83) and (36.45,117.84) .. (57.54,117.84) .. controls (78.64,117.84) and (95.75,133.83) .. (95.75,153.56) .. controls (95.75,173.28) and (78.64,189.27) .. (57.54,189.27) .. controls (36.45,189.27) and (19.34,173.28) .. (19.34,153.56) -- cycle ;
    \draw  [line width=1.5]  (38.09,153.56) .. controls (38.09,143.51) and (46.8,135.37) .. (57.54,135.37) .. controls (68.29,135.37) and (77,143.51) .. (77,153.56) .. controls (77,163.6) and (68.29,171.75) .. (57.54,171.75) .. controls (46.8,171.75) and (38.09,163.6) .. (38.09,153.56) -- cycle ;
    \draw  [fill={rgb, 255:red, 0; green, 0; blue, 0 }  ,fill opacity=1 ] (55.13,153.56) .. controls (55.13,152.31) and (56.21,151.3) .. (57.54,151.3) .. controls (58.88,151.3) and (59.96,152.31) .. (59.96,153.56) .. controls (59.96,154.8) and (58.88,155.81) .. (57.54,155.81) .. controls (56.21,155.81) and (55.13,154.8) .. (55.13,153.56) -- cycle ;
    \draw  [line width=1.5]  (138.29,132.66) .. controls (138.29,124.58) and (156.36,118.03) .. (178.65,118.03) .. controls (200.93,118.03) and (219,124.58) .. (219,132.66) .. controls (219,140.74) and (200.93,147.29) .. (178.65,147.29) .. controls (156.36,147.29) and (138.29,140.74) .. (138.29,132.66) -- cycle ;
    \draw  [line width=1.5]  (138.29,171.7) .. controls (138.29,163.62) and (156.36,157.07) .. (178.65,157.07) .. controls (200.93,157.07) and (219,163.62) .. (219,171.7) .. controls (219,179.78) and (200.93,186.33) .. (178.65,186.33) .. controls (156.36,186.33) and (138.29,179.78) .. (138.29,171.7) -- cycle ;
    \draw  [fill={rgb, 255:red, 0; green, 0; blue, 0 }  ,fill opacity=1 ] (82.02,127.98) .. controls (82.02,126.73) and (83.1,125.72) .. (84.43,125.72) .. controls (85.77,125.72) and (86.85,126.73) .. (86.85,127.98) .. controls (86.85,129.22) and (85.77,130.23) .. (84.43,130.23) .. controls (83.1,130.23) and (82.02,129.22) .. (82.02,127.98) -- cycle ;
    \draw  [fill={rgb, 255:red, 0; green, 0; blue, 0 }  ,fill opacity=1 ] (64.79,169.22) .. controls (64.79,167.98) and (65.87,166.97) .. (67.2,166.97) .. controls (68.53,166.97) and (69.62,167.98) .. (69.62,169.22) .. controls (69.62,170.47) and (68.53,171.48) .. (67.2,171.48) .. controls (65.87,171.48) and (64.79,170.47) .. (64.79,169.22) -- cycle ;
    \draw  [fill={rgb, 255:red, 0; green, 0; blue, 0 }  ,fill opacity=1 ] (150.86,144.42) .. controls (150.86,143.1) and (152.06,142.02) .. (153.54,142.02) .. controls (155.02,142.02) and (156.22,143.1) .. (156.22,144.42) .. controls (156.22,145.75) and (155.02,146.83) .. (153.54,146.83) .. controls (152.06,146.83) and (150.86,145.75) .. (150.86,144.42) -- cycle ;
    \draw  [fill={rgb, 255:red, 0; green, 0; blue, 0 }  ,fill opacity=1 ] (150.86,183.02) .. controls (150.86,181.69) and (152.06,180.61) .. (153.54,180.61) .. controls (155.02,180.61) and (156.22,181.69) .. (156.22,183.02) .. controls (156.22,184.35) and (155.02,185.42) .. (153.54,185.42) .. controls (152.06,185.42) and (150.86,184.35) .. (150.86,183.02) -- cycle ;
    \draw  [fill={rgb, 255:red, 0; green, 0; blue, 0 }  ,fill opacity=1 ] (176,152.46) .. controls (176,151.04) and (177.17,149.89) .. (178.61,149.89) .. controls (180.05,149.89) and (181.22,151.04) .. (181.22,152.46) .. controls (181.22,153.88) and (180.05,155.03) .. (178.61,155.03) .. controls (177.17,155.03) and (176,153.88) .. (176,152.46) -- cycle ;
    \draw  [dash pattern={on 4.5pt off 4.5pt}]  (153.54,144.42) -- (153.54,183.02) ;
    \draw  [dash pattern={on 4.5pt off 4.5pt}]  (178.65,132.66) -- (178.65,171.25) ;
    \draw  [dash pattern={on 0.84pt off 2.51pt}]  (153.54,144.42) -- (178.65,132.66) ;
    \draw  [dash pattern={on 0.84pt off 2.51pt}]  (153.54,183.02) -- (178.65,171.25) ;
    \draw  [line width=1.5]  (93.88,245.12) .. controls (93.17,243.44) and (92.52,241.75) .. (91.92,240.07) .. controls (84.29,218.72) and (88.46,205.94) .. (101.23,211.53) .. controls (114.01,217.12) and (130.54,238.96) .. (138.17,260.31) .. controls (138.18,260.33) and (138.19,260.35) .. (138.19,260.36) -- cycle ;
    \draw  [draw opacity=0][line width=1.5]  (111.93,218.88) .. controls (120.04,210.75) and (128.62,206.71) .. (135.34,208.9) .. controls (147.63,212.88) and (149.08,236.08) .. (138.59,260.72) .. controls (138.31,261.36) and (138.03,262) .. (137.75,262.64) -- (116.34,253.51) -- cycle ; \draw  [line width=1.5]  (111.93,218.88) .. controls (120.04,210.75) and (128.62,206.71) .. (135.34,208.9) .. controls (147.63,212.88) and (149.08,236.08) .. (138.59,260.72) .. controls (138.31,261.36) and (138.03,262) .. (137.75,262.64) ;
    \draw  [dash pattern={on 0.84pt off 2.51pt}]  (88.31,222.86) -- (145.53,239.63) ;
    \draw  [fill={rgb, 255:red, 0; green, 0; blue, 0 }  ,fill opacity=1 ] (86.31,222.86) .. controls (86.31,221.62) and (87.21,220.61) .. (88.31,220.61) .. controls (89.42,220.61) and (90.32,221.62) .. (90.32,222.86) .. controls (90.32,224.09) and (89.42,225.1) .. (88.31,225.1) .. controls (87.21,225.1) and (86.31,224.09) .. (86.31,222.86) -- cycle ;
    \draw  [fill={rgb, 255:red, 0; green, 0; blue, 0 }  ,fill opacity=1 ] (142.72,239.32) .. controls (142.72,238.08) and (143.62,237.08) .. (144.73,237.08) .. controls (145.83,237.08) and (146.73,238.08) .. (146.73,239.32) .. controls (146.73,240.56) and (145.83,241.57) .. (144.73,241.57) .. controls (143.62,241.57) and (142.72,240.56) .. (142.72,239.32) -- cycle ;
    \draw  [fill={rgb, 255:red, 0; green, 0; blue, 0 }  ,fill opacity=1 ] (112.17,251.88) .. controls (112.17,250.64) and (113.07,249.64) .. (114.17,249.64) .. controls (115.28,249.64) and (116.18,250.64) .. (116.18,251.88) .. controls (116.18,253.12) and (115.28,254.12) .. (114.17,254.12) .. controls (113.07,254.12) and (112.17,253.12) .. (112.17,251.88) -- cycle ;
    \draw  [dash pattern={on 4.5pt off 4.5pt}]  (123.64,233.18) -- (144.73,239.32) ;
    \draw  [draw opacity=0][line width=1.5]  (138.06,262.14) .. controls (137.16,263.72) and (136.2,265.29) .. (135.2,266.84) .. controls (121.24,288.26) and (102.48,297.4) .. (93.3,287.27) .. controls (86.16,279.37) and (86.93,262.11) .. (94.21,244.74) -- (118.58,248.48) -- cycle ; \draw  [line width=1.5]  (138.06,262.14) .. controls (137.16,263.72) and (136.2,265.29) .. (135.2,266.84) .. controls (121.24,288.26) and (102.48,297.4) .. (93.3,287.27) .. controls (86.16,279.37) and (86.93,262.11) .. (94.21,244.74) ;
    \draw  [draw opacity=0][line width=1.5]  (136.94,258.21) .. controls (145.42,278.3) and (146.19,294.61) .. (138.06,296.32) .. controls (133.27,297.32) and (126.33,293.05) .. (119.08,285.41) -- (116.19,253.15) -- cycle ; \draw  [color={rgb, 255:red, 0; green, 0; blue, 0 }  ,draw opacity=1 ][line width=1.5]  (136.94,258.21) .. controls (145.42,278.3) and (146.19,294.61) .. (138.06,296.32) .. controls (133.27,297.32) and (126.33,293.05) .. (119.08,285.41) ;
    \draw  [fill={rgb, 255:red, 0; green, 0; blue, 0 }  ,fill opacity=1 ] (79.85,21.15) -- (74.94,13.35) -- (83.85,15.88) -- (78.39,15.93) -- cycle ;
    \draw  [fill={rgb, 255:red, 0; green, 0; blue, 0 }  ,fill opacity=1 ] (199.89,73.57) -- (207.75,68.71) -- (205.19,77.55) -- (205.15,72.13) -- cycle ;
    \draw  [fill={rgb, 255:red, 0; green, 0; blue, 0 }  ,fill opacity=1 ] (76.42,126.24) -- (70.7,119.01) -- (79.82,120.58) -- (74.41,121.21) -- cycle ;
    \draw  [fill={rgb, 255:red, 0; green, 0; blue, 0 }  ,fill opacity=1 ] (203.8,24) -- (207.31,32.51) -- (198.97,28.53) -- (204.35,29.39) -- cycle ;
    \draw  [fill={rgb, 255:red, 0; green, 0; blue, 0 }  ,fill opacity=1 ] (71.52,162.28) -- (77.32,155.12) -- (77.84,164.3) -- (76,159.21) -- cycle ;
    \draw  [fill={rgb, 255:red, 0; green, 0; blue, 0 }  ,fill opacity=1 ] (142.19,230.1) -- (145.2,221.41) -- (148.83,229.87) -- (145.36,225.7) -- cycle ;
    \draw  [fill={rgb, 255:red, 0; green, 0; blue, 0 }  ,fill opacity=1 ] (203.11,179.84) -- (212.37,179.82) -- (205.5,185.99) -- (208.34,181.37) -- cycle ;
    \draw  [fill={rgb, 255:red, 0; green, 0; blue, 0 }  ,fill opacity=1 ] (206.11,140.35) -- (215.24,138.81) -- (209.5,146.02) -- (211.52,141) -- cycle ;
    \draw  [fill={rgb, 255:red, 0; green, 0; blue, 0 }  ,fill opacity=1 ] (84.76,212.99) -- (95.55,209.88) -- (91.91,216.21) -- (91.95,212.24) -- cycle ;

    \draw (53.11,52.93) node [anchor=north west][inner sep=0.75pt]  [font=\footnotesize] [align=left] {+2e};
    \draw (97.22,19.13) node [anchor=north west][inner sep=0.75pt]  [font=\footnotesize] [align=left] {\mbox{$-$}e};
    \draw (4.18,68) node [anchor=north west][inner sep=0.75pt]  [font=\footnotesize] [align=left] {\mbox{$-$}e};
    \draw (196.74,8.99) node [anchor=north west][inner sep=0.75pt]  [font=\footnotesize] [align=left] {\mbox{$-$}e};
    \draw (197.62,82.63) node [anchor=north west][inner sep=0.75pt]  [font=\footnotesize] [align=left] {\mbox{$-$}e};
    \draw (49.15,92.5) node [anchor=north west][inner sep=0.75pt]   [align=left] {(a)};
    \draw (167.71,94.45) node [anchor=north west][inner sep=0.75pt]   [align=left] {(b)};
    \draw (180.74,44.01) node [anchor=north west][inner sep=0.75pt]  [font=\footnotesize] [align=left] {+2e};
    \draw (48.61,189.4) node [anchor=north west][inner sep=0.75pt]   [align=left] {(c)};
    \draw (86.24,117.46) node [anchor=north west][inner sep=0.75pt]  [font=\footnotesize] [align=left] {\mbox{$-$}e};
    \draw (71.34,166.94) node [anchor=north west][inner sep=0.75pt]  [font=\footnotesize] [align=left] {\mbox{$-$}e};
    \draw (190.19,146.71) node [anchor=north west][inner sep=0.75pt]  [font=\footnotesize] [align=left] {+2e};
    \draw (131.55,140.31) node [anchor=north west][inner sep=0.75pt]  [font=\footnotesize] [align=left] {\mbox{$-$}e};
    \draw (132.13,180.86) node [anchor=north west][inner sep=0.75pt]  [font=\footnotesize] [align=left] {\mbox{$-$}e};
    \draw (169.62,192.97) node [anchor=north west][inner sep=0.75pt]   [align=left] {(d)};
    \draw (98.24,254.41) node [anchor=north west][inner sep=0.75pt]  [font=\footnotesize] [align=left] {+2e};
    \draw (68.81,213.75) node [anchor=north west][inner sep=0.75pt]  [font=\footnotesize] [align=left] {\mbox{$-$}e};
    \draw (149.36,231.34) node [anchor=north west][inner sep=0.75pt]  [font=\footnotesize] [align=left] {\mbox{$-$}e};
    \draw (111.17,295.69) node [anchor=north west][inner sep=0.75pt]   [align=left] {(e)};
    \draw (42.11,156.93) node [anchor=north west][inner sep=0.75pt]  [font=\footnotesize] [align=left] {+2e};
   \end{tikzpicture}
  \end{center}
\end{figure}

One aspect that deserves to be highlighted is the quest for symmetry (except in model (c) by Landé). Symmetry was seen as necessary to justify the stability of helium, for any unsymmetrical model---as the one of Landé---would let one of the electrons in a valence state, which was incoherent with helium's chemical properties. But symmetric models gave no convincing values for the ionization energy. The Bohr model (a), for instance, with a low electron shielding, gave a theoretical value for the first ionization energy of 28.8 eV \citetext{\citealt[p. 846]{RefWorks:van1922normal}}, while the accepted experimental value was, at the time, 25.4 $\pm$ 0.25 eV.

In the pursuit of a reasonable model under the Bohr assumptions, van Vleck calculated the ionization energy for the model with non-coplanar orbits proposed by Franck and Reiche. By then, every other proposed model had been discarded. One of the arguments against those models was provided by \citet{RefWorks:1449} on the basis of helium ionization by alpha particles. Millikan's observations "demand[ed] that both electrons find themselves in the same part of the atomic volume a very appreciable fraction of the time," a condition that could be satisfied by a model in which both electrons "are in orbits of much the same diameter but inclined to each other at angles of 60\textdegree or 90\textdegree" \citetext{\citealt[p. 459]{RefWorks:1449}}. The model of crossed orbital planes thus came to be the more reasonable option, so \citet[]{RefWorks:van1922normal} considered it suitable to tackle the calculation of ionization energies. Following tedious effort, however, the model gave an ionization potential of 20.7 eV, instead of the 25.4 eV demanded by experiment at the time.

Born considered that the situation demanded a more general analysis and jointly with Heisenberg \citetext{\citealt{RefWorks:1393}} tried to make some additional calculations to definitively set the acceptability of any possible model of helium under the quantum rules. \citet[]{RefWorks:1450}, in addition to a thorough review of the status of the helium model in the old quantum theory, discusses whether the authors were actually seeking to invalidate the existing models or were really just searching for a plausible, alternative model under the existing theories.

Their approach to the subject diverged from that of their predecessors. Instead of considering similar circular orbits for the electrons, they considered elliptic orbits and then they focused on the motion of their respective perihelia. They also considered the Rydberg correction as the experimental data to be fitted instead of the ionization energy. The Rydberg correction was, in effect, a measure of the shielding effect and thus also quantifies the interaction between the two electrons which, in turn, would determine the ionization energy (the lower the interaction, the higher the ionization energy). From this, they managed to get a general model depending on a single parameter whose values over a certain range determined the different combinations of rotational motion of the perihelia of both electrons. This allowed them to discard some of the combinations retaining the ones compatible with the established principles. Their \emph{modus operandi} included the use of perturbation theory, very well known to astronomers, which they used to calculate the Rydberg correction for those models.

Their Rydberg correction calculations disagreed, however, with the experimental results, thus consolidating the feeling of crisis of Bohr's atomic model and making them assert
\begin{quote}
  so we come to the conclusion that a consistent quantum-theoretical calculation of the helium problem leads to wrong values for the energy terms. There are obviously only two ways out of this difficulty: Either the quantum conditions are wrong [...] or the motion of the electrons no longer satisfies the mechanical equations. \citetext{\citealt[p. 243]{RefWorks:1393}}
\end{quote}

Heisenberg would work on the same problem in 1924, restricting the inner orbit to circular motion and achieving similarly unsuccessful results. His work was added as a section to \citet{RefWorks:1451}'s book \citetitle{RefWorks:1451}.

In 1925 three other major changes in atomic physics arose. The first was the paper by \citet[]{RefWorks:1452} who introduced the principle for the exclusion of "equivalent electrons". The second was Heisenberg's \citeyearpar{RefWorks:58} own paper on matrix mechanics. It is worth noticing that the recognition of the need of a fourth quantum number---associated with the magnetic properties of atoms and molecules---in the building scheme of the periodic table was interpreted by Heisenberg as a confirmation that the implicit discreteness of his matrix mechanics put it on the right track as a method to tackle the problems of atomic structure. The third was the suggestion by \comentario{Uhlenbeck and Goudsmit}\citet{RefWorks:uhlenbeck1925ersetzung} to consider the old "non-mechanical force" as the momentum associated with a spinning electron.

Shortly after Schrödinger's introduction of wave mechanics, Heisenberg returned to the problem of the multi-electron atom, this time with the tools of the new quantum mechanics. To this end, he worked on two papers, \citetitle{RefWorks:1264}, \citet{RefWorks:1264}, and \citetitle{RefWorks:1278}, \citet{RefWorks:1278}. In these papers he tried to give quantum-based arguments for a) the Bose--Einstein way of counting, b) a calculation of the Rydberg correction for helium and helium-like atoms, c) a justification of the non-combining terms of para- and ortho-helium, and d) a theoretical necessity for helium triplets.\footnote{At the time helium was considered to have just doublets.} In the next two sections we will enter in some detail into the papers.

On the one hand, we try here to outline the papers to make them available for historians and current physicists. Heisenberg was immersed in a very turbulent moment of physics and was thus using concepts and terminology not yet consolidated. So, trying to read the papers through present-day physical language and understanding can make it difficult to properly comprehend their meaning and historical relevance. The next section aims to bridge this gap.

On the other hand we attempt to connect with the first claim expressed in the \nameref{introduction}: mainly a) that indistinguishability played a role in Heisenberg's mathematical formulation of the symmetrization postulate and subsequently the Pauli exclusion principle and the interpretation of the double spectrum of helium (para- and ortho-helium), and b) that Heisenberg's new way of using superposition was a relevant feature of his proposal for the dynamics of the helium atom.

\section{Heisenberg's new approach to the helium problem}\label{1926}

\subsection{Matrices and wave functions}\label{1926a}
Heisenberg's first paper, \citetitle{RefWorks:1264}, begins with a general review of the recent changes (de Broglie, Bose, Schrödinger, Pauli...) to conclude that a particle-structure view of matter fits nature better than the wave-structure view, even accepting that the motion of particles could not be governed by the usual spacetime considerations. Then, Heisenberg states the difficulties of accommodating Bose--Einstein statistics and Pauli's exclusion principle in quantum mechanics but announces "in advance", as he put it, the result that these problems are solved and that there is a natural relationship between Bose--Einstein statistics and quantum mechanics.\footnote{The comment at sect. \ref{ac}-\nameref{ac} applies to this section.}

His starting point is the classical problem of resonance in which two coupled linear harmonic oscillators formed by two charged particles oscillate around their respective zero points in one of the proper modes of coupling. Next, he considers the energy quantized in half-integer form as needed for the application of quantum rules,\footnote{$n_1^{\prime}$ and $n_2^{\prime}$ are integers, and $\omega_1^{\prime}$, $\omega_2^{\prime}$ are the frequencies of the oscillators.}
\begin{equation}\label{H5}
H_{n_1^{\prime}, n_2^{\prime}}=\frac{\omega_1^{\prime} h}{2 \pi}\left(n_1^{\prime}+\frac{1}{2}\right)+\frac{\omega_2^{\prime} h}{2 \pi}\left(n_2^{\prime}+\frac{1}{2}\right). 
\end{equation}
The energy $H$ of this model system corresponds, by analogy, to the perturbation energy of the helium problem. This perturbation energy is understood to result from the (repulsive) interaction between the two electrons which is added to the otherwise hydrogen-like electrons orbiting around the nucleus. This energy will reduce the attraction of the electron by the nucleus, and is the key element of Heisenberg's calculation. Then, he begins an argument dealing simultaneously with three different physical problems. The first of these is the classical Hamiltonian for the coupled harmonic oscillator,
\begin{equation}
H=\frac{1}{2 m} p_1^{\prime 2}+\frac{m}{2} \omega^{\prime 2} q_1^{\prime 2}+\frac{1}{2 m} p_2^{\prime 2}+\frac{m}{2} \omega^{\prime 2} q_2^{\prime 2}. 
\end{equation}
Second is the quantum Hamiltonian \eqref{H5} for the same problem; third, the multi-body atomic problem. Starting from the classical Hamiltonian and combining the results with the quantum Hamiltonian, he defines two sets of energy levels which labels as the set of (\textbullet) and the set of (+) (see figure \ref{HFig2} at p. \pageref{HFig2}).

\begin{figure}[htp]
  \caption{Heisenberg's first (classical) approach} \label{HFig2}
  \begin{center}
  \tikzset{every picture/.style={line width=0.75pt}} 

  \begin{tikzpicture}[x=0.75pt,y=0.75pt,yscale=-1,xscale=1]

  \draw  [fill={rgb, 255:red, 0; green, 0; blue, 0 }  ,fill opacity=1 ] (186.92,10.58) .. controls (186.92,8.6) and (188.52,7) .. (190.5,7) .. controls (192.48,7) and (194.08,8.6) .. (194.08,10.58) .. controls (194.08,12.56) and (192.48,14.17) .. (190.5,14.17) .. controls (188.52,14.17) and (186.92,12.56) .. (186.92,10.58) -- cycle ;
  \draw  [line width=1.5]  (226.42,25.13) -- (238.08,25.13)(232.25,19) -- (232.25,31.25) ;
  \draw  [fill={rgb, 255:red, 0; green, 0; blue, 0 }  ,fill opacity=1 ] (186.92,67.58) .. controls (186.92,65.6) and (188.52,64) .. (190.5,64) .. controls (192.48,64) and (194.08,65.6) .. (194.08,67.58) .. controls (194.08,69.56) and (192.48,71.17) .. (190.5,71.17) .. controls (188.52,71.17) and (186.92,69.56) .. (186.92,67.58) -- cycle ;
  \draw  [fill={rgb, 255:red, 0; green, 0; blue, 0 }  ,fill opacity=1 ] (186.92,140.58) .. controls (186.92,138.6) and (188.52,137) .. (190.5,137) .. controls (192.48,137) and (194.08,138.6) .. (194.08,140.58) .. controls (194.08,142.56) and (192.48,144.17) .. (190.5,144.17) .. controls (188.52,144.17) and (186.92,142.56) .. (186.92,140.58) -- cycle ;
  \draw  [fill={rgb, 255:red, 0; green, 0; blue, 0 }  ,fill opacity=1 ] (186.92,206.58) .. controls (186.92,204.6) and (188.52,203) .. (190.5,203) .. controls (192.48,203) and (194.08,204.6) .. (194.08,206.58) .. controls (194.08,208.56) and (192.48,210.17) .. (190.5,210.17) .. controls (188.52,210.17) and (186.92,208.56) .. (186.92,206.58) -- cycle ;
  \draw  [fill={rgb, 255:red, 0; green, 0; blue, 0 }  ,fill opacity=1 ] (275.92,38.58) .. controls (275.92,36.6) and (277.52,35) .. (279.5,35) .. controls (281.48,35) and (283.08,36.6) .. (283.08,38.58) .. controls (283.08,40.56) and (281.48,42.17) .. (279.5,42.17) .. controls (277.52,42.17) and (275.92,40.56) .. (275.92,38.58) -- cycle ;
  \draw  [fill={rgb, 255:red, 0; green, 0; blue, 0 }  ,fill opacity=1 ] (275.92,104.58) .. controls (275.92,102.6) and (277.52,101) .. (279.5,101) .. controls (281.48,101) and (283.08,102.6) .. (283.08,104.58) .. controls (283.08,106.56) and (281.48,108.17) .. (279.5,108.17) .. controls (277.52,108.17) and (275.92,106.56) .. (275.92,104.58) -- cycle ;
  \draw  [line width=1.5]  (226.42,84.13) -- (238.08,84.13)(232.25,78) -- (232.25,90.25) ;
  \draw  [line width=1.5]  (226.42,147.13) -- (238.08,147.13)(232.25,141) -- (232.25,153.25) ;
  \draw  [line width=1.5]  (314.67,55.13) -- (326.33,55.13)(320.5,49) -- (320.5,61.25) ;

  \draw (182,212) node [anchor=north west][inner sep=0.75pt]  [font=\footnotesize] [align=left] {{\fontfamily{ptm}\selectfont 0,0}};
  \draw (182,146.34) node [anchor=north west][inner sep=0.75pt]  [font=\footnotesize] [align=left] {{\fontfamily{ptm}\selectfont 1,0}};
  \draw (182,73.67) node [anchor=north west][inner sep=0.75pt]  [font=\footnotesize] [align=left] {{\fontfamily{ptm}\selectfont 2,0}};
  \draw (182,17) node [anchor=north west][inner sep=0.75pt]  [font=\footnotesize] [align=left] {{\fontfamily{ptm}\selectfont 3,0}};
  \draw (223.75,157) node [anchor=north west][inner sep=0.75pt]  [font=\footnotesize] [align=left] {{\fontfamily{ptm}\selectfont 0,1}};
  \draw (223.75,93) node [anchor=north west][inner sep=0.75pt]  [font=\footnotesize] [align=left] {{\fontfamily{ptm}\selectfont 1,1}};
  \draw (223.75,34) node [anchor=north west][inner sep=0.75pt]  [font=\footnotesize] [align=left] {{\fontfamily{ptm}\selectfont 2,1}};
  \draw (271,111) node [anchor=north west][inner sep=0.75pt]  [font=\footnotesize] [align=left] {{\fontfamily{ptm}\selectfont 0,2}};
  \draw (269,44) node [anchor=north west][inner sep=0.75pt]  [font=\footnotesize] [align=left] {{\fontfamily{ptm}\selectfont 1,2}};
  \draw (312,64) node [anchor=north west][inner sep=0.75pt]  [font=\footnotesize] [align=left] {{\fontfamily{ptm}\selectfont 0,3}};
  \end{tikzpicture}
  \end{center}
\end{figure}

Heisenberg next extends the problem of the linear harmonic oscillator to the bound electron around a nucleus by considering, in addition to dipole momenta, higher-order electric momenta (quadrupole or higher-order poles) assuming the non-spherical symmetry of the nucleus and the shielding of the surrounding electrons. Working on an argument based on the even degree of the polynomial arising from calculating the energy due to the quadrupole, he concludes that the two systems proposed (\textbullet, +) have no transitions between them to the extent that the physical characteristics (mass, frequency) of the oscillating particles remain the same. A comment on this: at first reading, it may appear that Heisenberg was drawing conclusions for the quantum Hamiltonian from the classical Hamiltonian. It is just the opposite. To give an analogy of the quantum behavior, Heisenberg was assuming that the two classical oscillators were quantized and could only vibrate at integer multiples of a certain energy. This can be better appreciated in Birtwistle's textbook when the latter uses the classical solution of the oscillator with a fixed frequency and links any other possible frequency to an integer $\tau$ and "the corresponding quantum number[s]" $n_1'$ and $n_2'$ of \eqref{H5} \citetext{\citealt[pp. 217-218]{RefWorks:1270}}.

Then, Heisenberg claims the indefiniteness of the quantum mechanical solutions to be the most relevant outcome of his work arguing that it allows a common frame for the Bose--Einstein statistics and the Pauli exclusion principle. As it will be argued in section \ref{discussion}, the claimed integration of Bose--Einstein counting into the common frame of Heisenberg's matrix mechanics was not really accomplished in this work. The indefiniteness that Heisenberg refers to rests on the assumption that the permutation of the electron wave functions provides a new solution for the wave equation (indirectly leading to a form of the Pauli exclusion principle); it is, though, hardly recognizable in this first paper. More on this later at section \ref{Historicalsummary}.

Heisenberg then moves to his matrix mechanics, introducing, for the case of two identical systems (electrons $a$ and $b$), the interaction energy $H_1$ in such a way that it depends on the stationary states of the identical systems. He remarks that the classical concept of resonance differs from the quantum mechanical one in that classical resonance depends on the natural frequencies of both systems being equal, while quantum mechanical resonance depends not on the stationary frequencies but on the emission frequency of one system being equal to the absorption frequency of the other. He sees degeneracy from two different points of view. On the one hand, he considers a simplified model of the whole atom without electron--electron interaction, so that the exchange of the identical electrons does not change their energies; this would correspond to a (new) quantum mechanical concept of degeneracy. On the other hand, he considers also interaction between the electrons, stating that degeneracy is then removed if this interaction is assumed. This second case would correspond to the classical concept of degeneracy: although both electrons are initially assumed to oscillate with the same natural frequency, the coupling, which forces the energy back and forth between them, removes this degeneracy.

Via the transformation matrices he had already used in the "three-men-paper" \citetext{\citealt[]{RefWorks:60}}, Heisenberg builds a system of equations based on a set of symmetries that allows him to find a transformation matrix $\mathbf S$ in such a way that the energy matrix $\mathbf{W}$ can be written in diagonal form
\begin{equation}\label{0324}
\begin{aligned}
\mathbf{W}&=
\left(
\begin{array}{ll}
W_1 &                               \\
                              & W_2
\end{array}
\right)                             \\
          &=
\left(
\begin{array}{ll}
H^1(n m, n m) + H^1(n m, m n) &                               \\
                              & H^1(n m, n m) - H^1(n m, m n)
\end{array}
\right),
\end{aligned}
\end{equation}
with
\begin{equation*}
\mathbf{S}=
\left(
\begin{array}{cc}
\frac{1}{\sqrt{2}} & \frac{1}{\sqrt{2}} \\
\frac{1}{\sqrt{2}} & -\frac{1}{\sqrt{2}}
\end{array}
\right),
\end{equation*}
so that he can eventually write
\begin{equation}\label{0353}
\mathbf{f}^{\prime}=\mathbf{S}^{-1} \mathbf{f} \mathbf{S}.  
\end{equation}
At eq. \eqref{0353}, $\mathbf{f}$ ($f_{n_1 m_1, n_2 m_2}$) means a four-indexed matrix representing the amplitudes of the transitions associated with the electrons in combination with their energy levels. The first and second subscripts, respectively, represent the energy levels of electrons $a$ and $b$ prior to the transition. The third and fourth subscripts represent the energy levels of $a$ and $b$ after the transition, and $\mathbf{S}$ is a unitary matrix calculated from basic assumptions of symmetry. The matrix elements ($f_{n_1 m_1, n_2 m_2}$) are, then, the amplitudes corresponding to the transitions so that $f_{n_1 m_1, n_2 m_2}=0$ means no transition for the joint system $(a,b)$ between levels $(n_1 m_1)$ and $(n_2 m_2)$.

Our meaning is that the set of symmetries used to state expression \eqref{0324},
\begin{equation}\label{0354}
H^1(n m, n m)=H^1(m n, m n), \quad H^1(n m, m n)=H^1(m n, n m). 
\end{equation}
assumes that energy does not change in case of particle exchange, a consideration that Heisenberg perhaps borrowed from statistical mechanics.\footnote{It is more clearly seen in the second expression of \eqref{0354} which states that the energy does not change if instead of electron $a$ changing between levels $n \rightarrow m$, and $b$ between levels $m \rightarrow n$, one swaps the electrons (or the levels they are in).} This property is present in his subsequent use of Schrödinger's wave functions and is relevant to understanding Heisenberg's claim of accommodating Pauli's principle and Bose--Einstein statistics within quantum mechanics.

Next, Heisenberg proceeds to justifying the non-combination (no spectral lines) between the systems representing para- and ortho-helium (resp. \textbullet\;  and  +). To this end, he builds subscripts representing different elements of both systems (\textbullet\; and +) and checks that applying the matrix $\mathbf{S}$ in the corresponding form results in a zero probability of transition between them. Heisenberg's use of the matrix subscripts in this formulation is, though essentially correct, somewhat tricky and seems misleading in a first reading. An attempt at clarification has been made at \citet[vol. 3, p. 293]{RefWorks:40} by slightly modifying Heisenberg's assumptions but it does not really unravel Heisenberg's train of thought. Although a correct explanation is possible, it is lengthy so we will mention just a few aspects here. In the first place, Heisenberg proceeds to change the old subscripts of his initial scheme (Fig. \ref{HFig2}), adapting them (the first index is changed from 0 to 1) to better match the matrix representation used in his calculations to obtain a new schematic image of the spectrum (Fig. \ref{HFig4} in p. \pageref{HFig4}).
\begin{figure}[htp]
  \caption{Heisenberg's Fig. 4. An interpretation of Heisenberg's approach under quantum (matrix) considerations} \label{HFig4}
  \begin{center}
  \tikzset{every picture/.style={line width=0.75pt}} 
  \begin{tikzpicture}[x=0.75pt,y=0.75pt,yscale=-1,xscale=1]

  \draw  [fill={rgb, 255:red, 0; green, 0; blue, 0 }  ,fill opacity=1 ] (176.73,79.27) .. controls (176.73,76.91) and (178.64,75) .. (181,75) .. controls (183.36,75) and (185.27,76.91) .. (185.27,79.27) .. controls (185.27,81.64) and (183.36,83.55) .. (181,83.55) .. controls (178.64,83.55) and (176.73,81.64) .. (176.73,79.27) -- cycle ;
  \draw  [line width=1.5]  (238.73,87.63) -- (252.27,87.63)(245.5,81) -- (245.5,94.25) ;
  \draw  [fill={rgb, 255:red, 0; green, 0; blue, 0 }  ,fill opacity=1 ] (176.73,112.27) .. controls (176.73,109.91) and (178.64,108) .. (181,108) .. controls (183.36,108) and (185.27,109.91) .. (185.27,112.27) .. controls (185.27,114.64) and (183.36,116.55) .. (181,116.55) .. controls (178.64,116.55) and (176.73,114.64) .. (176.73,112.27) -- cycle ;
  \draw  [fill={rgb, 255:red, 0; green, 0; blue, 0 }  ,fill opacity=1 ] (176.73,163.27) .. controls (176.73,160.91) and (178.64,159) .. (181,159) .. controls (183.36,159) and (185.27,160.91) .. (185.27,163.27) .. controls (185.27,165.64) and (183.36,167.55) .. (181,167.55) .. controls (178.64,167.55) and (176.73,165.64) .. (176.73,163.27) -- cycle ;
  \draw  [fill={rgb, 255:red, 0; green, 0; blue, 0 }  ,fill opacity=1 ] (176.73,232.27) .. controls (176.73,229.91) and (178.64,228) .. (181,228) .. controls (183.36,228) and (185.27,229.91) .. (185.27,232.27) .. controls (185.27,234.64) and (183.36,236.55) .. (181,236.55) .. controls (178.64,236.55) and (176.73,234.64) .. (176.73,232.27) -- cycle ;
  \draw  [fill={rgb, 255:red, 0; green, 0; blue, 0 }  ,fill opacity=1 ] (176.73,303.27) .. controls (176.73,300.91) and (178.64,299) .. (181,299) .. controls (183.36,299) and (185.27,300.91) .. (185.27,303.27) .. controls (185.27,305.64) and (183.36,307.55) .. (181,307.55) .. controls (178.64,307.55) and (176.73,305.64) .. (176.73,303.27) -- cycle ;
  \draw  [fill={rgb, 255:red, 0; green, 0; blue, 0 }  ,fill opacity=1 ] (207,379.27) .. controls (207,376.91) and (208.91,375) .. (211.27,375) .. controls (213.64,375) and (215.55,376.91) .. (215.55,379.27) .. controls (215.55,381.64) and (213.64,383.55) .. (211.27,383.55) .. controls (208.91,383.55) and (207,381.64) .. (207,379.27) -- cycle ;
  \draw  [fill={rgb, 255:red, 0; green, 0; blue, 0 }  ,fill opacity=1 ] (345.73,66.27) .. controls (345.73,63.91) and (347.64,62) .. (350,62) .. controls (352.36,62) and (354.27,63.91) .. (354.27,66.27) .. controls (354.27,68.64) and (352.36,70.55) .. (350,70.55) .. controls (347.64,70.55) and (345.73,68.64) .. (345.73,66.27) -- cycle ;
  \draw  [fill={rgb, 255:red, 0; green, 0; blue, 0 }  ,fill opacity=1 ] (345.73,130.27) .. controls (345.73,127.91) and (347.64,126) .. (350,126) .. controls (352.36,126) and (354.27,127.91) .. (354.27,130.27) .. controls (354.27,132.64) and (352.36,134.55) .. (350,134.55) .. controls (347.64,134.55) and (345.73,132.64) .. (345.73,130.27) -- cycle ;
  \draw  [fill={rgb, 255:red, 0; green, 0; blue, 0 }  ,fill opacity=1 ] (374,203.27) .. controls (374,200.91) and (375.91,199) .. (378.27,199) .. controls (380.64,199) and (382.55,200.91) .. (382.55,203.27) .. controls (382.55,205.64) and (380.64,207.55) .. (378.27,207.55) .. controls (375.91,207.55) and (374,205.64) .. (374,203.27) -- cycle ;
  \draw  [line width=1.5]  (238.73,121.63) -- (252.27,121.63)(245.5,115) -- (245.5,128.25) ;
  \draw  [line width=1.5]  (238.73,175.63) -- (252.27,175.63)(245.5,169) -- (245.5,182.25) ;
  \draw  [line width=1.5]  (238.73,244.63) -- (252.27,244.63)(245.5,238) -- (245.5,251.25) ;
  \draw  [line width=1.5]  (238.73,312.63) -- (252.27,312.63)(245.5,306) -- (245.5,319.25) ;
  \draw  [line width=1.5]  (400.73,78.63) -- (414.27,78.63)(407.5,72) -- (407.5,85.25) ;
  \draw  [line width=1.5]  (400.73,136.63) -- (414.27,136.63)(407.5,130) -- (407.5,143.25) ;

  \draw (172.5,240) node [anchor=north west][inner sep=0.75pt]  [font=\footnotesize] [align=left] {{\fontfamily{ptm}\selectfont 3,1}};
  \draw (172.5,169.34) node [anchor=north west][inner sep=0.75pt]  [font=\footnotesize] [align=left] {{\fontfamily{ptm}\selectfont 4,1}};
  \draw (172.5,118.67) node [anchor=north west][inner sep=0.75pt]  [font=\footnotesize] [align=left] {{\fontfamily{ptm}\selectfont 5,1}};
  \draw (172.5,86) node [anchor=north west][inner sep=0.75pt]  [font=\footnotesize] [align=left] {{\fontfamily{ptm}\selectfont 6,1}};
  \draw (237,185) node [anchor=north west][inner sep=0.75pt]  [font=\footnotesize] [align=left] {{\fontfamily{ptm}\selectfont 1,4}};
  \draw (237,130) node [anchor=north west][inner sep=0.75pt]  [font=\footnotesize] [align=left] {{\fontfamily{ptm}\selectfont 1,5}};
  \draw (237,96) node [anchor=north west][inner sep=0.75pt]  [font=\footnotesize] [align=left] {{\fontfamily{ptm}\selectfont 1,6}};
  \draw (341.5,137) node [anchor=north west][inner sep=0.75pt]  [font=\footnotesize] [align=left] {{\fontfamily{ptm}\selectfont 3,2}};
  \draw (341.5,73) node [anchor=north west][inner sep=0.75pt]  [font=\footnotesize] [align=left] {{\fontfamily{ptm}\selectfont 4,2}};
  \draw (399,87) node [anchor=north west][inner sep=0.75pt]  [font=\footnotesize] [align=left] {{\fontfamily{ptm}\selectfont 2,4}};
  \draw (167,36.5) node [anchor=north west][inner sep=0.75pt]   [align=left] {{\fontfamily{ptm}\selectfont Para-}};
  \draw (227,36.5) node [anchor=north west][inner sep=0.75pt]   [align=left] {{\fontfamily{ptm}\selectfont Ortho-}};
  \draw (336,36.5) node [anchor=north west][inner sep=0.75pt]   [align=left] {{\fontfamily{ptm}\selectfont Para-}};
  \draw (389,36.5) node [anchor=north west][inner sep=0.75pt]   [align=left] {{\fontfamily{ptm}\selectfont Ortho-}};
  \draw (172.5,309) node [anchor=north west][inner sep=0.75pt]  [font=\footnotesize] [align=left] {{\fontfamily{ptm}\selectfont 2,1}};
  \draw (203,385) node [anchor=north west][inner sep=0.75pt]  [font=\footnotesize] [align=left] {{\fontfamily{ptm}\selectfont 1,1}};
  \draw (237,253) node [anchor=north west][inner sep=0.75pt]  [font=\footnotesize] [align=left] {{\fontfamily{ptm}\selectfont 1,3}};
  \draw (237,321) node [anchor=north west][inner sep=0.75pt]  [font=\footnotesize] [align=left] {{\fontfamily{ptm}\selectfont 1,2}};
  \draw (370,210) node [anchor=north west][inner sep=0.75pt]  [font=\footnotesize] [align=left] {{\fontfamily{ptm}\selectfont 2,2}};
  \draw (399,145) node [anchor=north west][inner sep=0.75pt]  [font=\footnotesize] [align=left] {{\fontfamily{ptm}\selectfont 2,3}};
  \end{tikzpicture}
  \end{center}
\end{figure}

Notice that, once the classical scheme has been dismissed, there is no further need for any argument related to the subscripts; each single subscript now represents the set of quantum numbers that determine the (spatial) state functions and do not stand for the energy levels of a quantum or classical oscillator as before. Second, Heisenberg assumes that the system \textbullet \, corresponds to the terms of para-helium with a symmetric (spatial) state function and the system + corresponds to the terms of ortho-helium with an antisymmetric (spatial) state function. It is worth emphasizing that the matrix elements that Heisenberg has calculated hitherto correspond, strictly speaking, to state \emph{functions without spin}, so they would correspond to today's "spatial state functions". At that time, such specific terminology had not yet evolved.

Heisenberg then re-states the same concept, not from matrix mechanics, but from Schrödinger's formulation where he can deal with the (spatial) state functions in a more specific way, thus anticipating what he will later do in the second paper, \citet[]{RefWorks:1278}, on the helium problem. In this formulation, the absence of transitions between para- and ortho-helium is connected with the following integral being zero ($f$ represents an arbitrary symmetrical operator)
 \begin{equation}\label{0233}
\int f \frac{1}{2}\left(\varphi_n^a \varphi_m^b+\varphi_m^a \varphi_n^b\right) \overline{\left(\varphi_n^a\right.} \overline{\varphi_m^b}-\overline{\varphi_m^a} \overline{\varphi_n^b})\, d q_1^a \ldots d q_f^b, 
\end{equation}
---a fact that Heisenberg uses to illustrate the effects of the permutation invariance of the electrons. Some of the terms appearing in this integral play also a role, by means of the Schrödinger functions, in the calculation of interaction energies made in the paper on helium, \citet[]{RefWorks:1278}, where the terms are embedded in the statistical factors provided by the Schrödinger wave functions. This gives a physical sense to Heisenberg's claim that "the integral retains its value" under exchange of the systems [the electrons] \citetext{\citealt[p. 420]{RefWorks:1264}}. Indeed, notice that the integral of Schrödinger functions (used by Heisenberg to calculate the Rydberg correction for helium) stands for an average of a physical quantity (the interaction energy). It is in this averaging process that it makes sense to relate Heisenberg's arguments on electron exchange to statistical mechanics. This general method is what Heisenberg uses in his second paper and what allows him to connect with the above discussed matrix approach presented in the first section of his first paper, \citetitle[]{RefWorks:1264}.

Next, Heisenberg sketches the physical assumptions underlying his claims and announces that a quantitative treatment will be offered later (in the paper on the helium atom). He distinguishes between two levels of accuracy: the case where spin is ignored and the case where spin is considered. When ignoring the spin, a simple Coulomb repulsion has to be considered that was not present in hydrogen-like atoms. The existence of two non-combining spectra (for para- and ortho-helium) can be theoretically justified via an explanation that mixes Heisenberg's matrix mechanics (which allows an explanation based on the symmetry of the terms involved in the calculation) with Schrödinger's wave mechanics (which will allow the effective calculation of energy averages already announced by Heisenberg). The appearance of the ground state in just the para-helium system is taken by Heisenberg as a confirmation of the soundness of the theory. Then, he advances a picture "that corresponds somewhat to the quantum mechanical solution": one must imagine that the electrons are periodically exchanging places continuously with constant period in analogy to the periodic energy swings seen in the mentioned example of the coupled oscillators, where the period of the swinging is given by the difference between para-helium and ortho-helium \citetext{\citealt[p. 421]{RefWorks:1264}}.

\begin{figure}[htp]
  \caption{Heisenberg's Fig. 5. Para- and ortho-helium scheme considering spin. Now the \textbullet \, (resp. +) represent antisymmetric (resp. symmetric) \emph{global} state functions} \label{HFig5}
  \begin{center}
  \tikzset{every picture/.style={line width=0.75pt}} 

  \begin{tikzpicture}[x=0.75pt,y=0.75pt,yscale=-1,xscale=1]

  \draw  [fill={rgb, 255:red, 0; green, 0; blue, 0 }  ,fill opacity=1 ] (97.73,79.27) .. controls (97.73,76.91) and (99.64,75) .. (102,75) .. controls (104.36,75) and (106.27,76.91) .. (106.27,79.27) .. controls (106.27,81.64) and (104.36,83.55) .. (102,83.55) .. controls (99.64,83.55) and (97.73,81.64) .. (97.73,79.27) -- cycle ;
  \draw  [line width=1.5]  (289.73,90.63) -- (303.27,90.63)(296.5,84) -- (296.5,97.25) ;
  \draw  [fill={rgb, 255:red, 0; green, 0; blue, 0 }  ,fill opacity=1 ] (97.73,145.77) .. controls (97.73,143.41) and (99.64,141.5) .. (102,141.5) .. controls (104.36,141.5) and (106.27,143.41) .. (106.27,145.77) .. controls (106.27,148.14) and (104.36,150.05) .. (102,150.05) .. controls (99.64,150.05) and (97.73,148.14) .. (97.73,145.77) -- cycle ;
  \draw  [fill={rgb, 255:red, 0; green, 0; blue, 0 }  ,fill opacity=1 ] (97.73,212.27) .. controls (97.73,209.91) and (99.64,208) .. (102,208) .. controls (104.36,208) and (106.27,209.91) .. (106.27,212.27) .. controls (106.27,214.64) and (104.36,216.55) .. (102,216.55) .. controls (99.64,216.55) and (97.73,214.64) .. (97.73,212.27) -- cycle ;
  \draw  [fill={rgb, 255:red, 0; green, 0; blue, 0 }  ,fill opacity=1 ] (207,302.27) .. controls (207,299.91) and (208.91,298) .. (211.27,298) .. controls (213.64,298) and (215.55,299.91) .. (215.55,302.27) .. controls (215.55,304.64) and (213.64,306.55) .. (211.27,306.55) .. controls (208.91,306.55) and (207,304.64) .. (207,302.27) -- cycle ;
  \draw  [line width=1.5]  (289.73,158.13) -- (303.27,158.13)(296.5,151.5) -- (296.5,164.75) ;
  \draw  [line width=1.5]  (289.73,225.63) -- (303.27,225.63)(296.5,219) -- (296.5,232.25) ;
  \draw  [line width=1.5]  (109.73,70.63) -- (123.27,70.63)(116.5,64) -- (116.5,77.25) ;
  \draw  [line width=1.5]  (122.73,78.63) -- (136.27,78.63)(129.5,72) -- (129.5,85.25) ;
  \draw  [line width=1.5]  (134.73,86.63) -- (148.27,86.63)(141.5,80) -- (141.5,93.25) ;
  \draw  [line width=1.5]  (106.73,133.63) -- (120.27,133.63)(113.5,127) -- (113.5,140.25) ;
  \draw  [line width=1.5]  (119.73,141.63) -- (133.27,141.63)(126.5,135) -- (126.5,148.25) ;
  \draw  [line width=1.5]  (131.73,149.63) -- (145.27,149.63)(138.5,143) -- (138.5,156.25) ;
  \draw  [line width=1.5]  (108.73,202.63) -- (122.27,202.63)(115.5,196) -- (115.5,209.25) ;
  \draw  [line width=1.5]  (121.73,210.63) -- (135.27,210.63)(128.5,204) -- (128.5,217.25) ;
  \draw  [line width=1.5]  (133.73,218.63) -- (147.27,218.63)(140.5,212) -- (140.5,225.25) ;
  \draw  [line width=1.5]  (214.73,291.63) -- (228.27,291.63)(221.5,285) -- (221.5,298.25) ;
  \draw  [line width=1.5]  (227.73,299.63) -- (241.27,299.63)(234.5,293) -- (234.5,306.25) ;
  \draw  [line width=1.5]  (239.73,307.63) -- (253.27,307.63)(246.5,301) -- (246.5,314.25) ;
  \draw  [fill={rgb, 255:red, 0; green, 0; blue, 0 }  ,fill opacity=1 ] (301,78.27) .. controls (301,75.91) and (302.91,74) .. (305.27,74) .. controls (307.64,74) and (309.55,75.91) .. (309.55,78.27) .. controls (309.55,80.64) and (307.64,82.55) .. (305.27,82.55) .. controls (302.91,82.55) and (301,80.64) .. (301,78.27) -- cycle ;
  \draw  [fill={rgb, 255:red, 0; green, 0; blue, 0 }  ,fill opacity=1 ] (311,87.27) .. controls (311,84.91) and (312.91,83) .. (315.27,83) .. controls (317.64,83) and (319.55,84.91) .. (319.55,87.27) .. controls (319.55,89.64) and (317.64,91.55) .. (315.27,91.55) .. controls (312.91,91.55) and (311,89.64) .. (311,87.27) -- cycle ;
  \draw  [fill={rgb, 255:red, 0; green, 0; blue, 0 }  ,fill opacity=1 ] (320,96.27) .. controls (320,93.91) and (321.91,92) .. (324.27,92) .. controls (326.64,92) and (328.55,93.91) .. (328.55,96.27) .. controls (328.55,98.64) and (326.64,100.55) .. (324.27,100.55) .. controls (321.91,100.55) and (320,98.64) .. (320,96.27) -- cycle ;
  \draw  [fill={rgb, 255:red, 0; green, 0; blue, 0 }  ,fill opacity=1 ] (302,147.27) .. controls (302,144.91) and (303.91,143) .. (306.27,143) .. controls (308.64,143) and (310.55,144.91) .. (310.55,147.27) .. controls (310.55,149.64) and (308.64,151.55) .. (306.27,151.55) .. controls (303.91,151.55) and (302,149.64) .. (302,147.27) -- cycle ;
  \draw  [fill={rgb, 255:red, 0; green, 0; blue, 0 }  ,fill opacity=1 ] (312,156.27) .. controls (312,153.91) and (313.91,152) .. (316.27,152) .. controls (318.64,152) and (320.55,153.91) .. (320.55,156.27) .. controls (320.55,158.64) and (318.64,160.55) .. (316.27,160.55) .. controls (313.91,160.55) and (312,158.64) .. (312,156.27) -- cycle ;
  \draw  [fill={rgb, 255:red, 0; green, 0; blue, 0 }  ,fill opacity=1 ] (321,165.27) .. controls (321,162.91) and (322.91,161) .. (325.27,161) .. controls (327.64,161) and (329.55,162.91) .. (329.55,165.27) .. controls (329.55,167.64) and (327.64,169.55) .. (325.27,169.55) .. controls (322.91,169.55) and (321,167.64) .. (321,165.27) -- cycle ;
  \draw  [fill={rgb, 255:red, 0; green, 0; blue, 0 }  ,fill opacity=1 ] (302,215.27) .. controls (302,212.91) and (303.91,211) .. (306.27,211) .. controls (308.64,211) and (310.55,212.91) .. (310.55,215.27) .. controls (310.55,217.64) and (308.64,219.55) .. (306.27,219.55) .. controls (303.91,219.55) and (302,217.64) .. (302,215.27) -- cycle ;
  \draw  [fill={rgb, 255:red, 0; green, 0; blue, 0 }  ,fill opacity=1 ] (312,224.27) .. controls (312,221.91) and (313.91,220) .. (316.27,220) .. controls (318.64,220) and (320.55,221.91) .. (320.55,224.27) .. controls (320.55,226.64) and (318.64,228.55) .. (316.27,228.55) .. controls (313.91,228.55) and (312,226.64) .. (312,224.27) -- cycle ;
  \draw  [fill={rgb, 255:red, 0; green, 0; blue, 0 }  ,fill opacity=1 ] (321,233.27) .. controls (321,230.91) and (322.91,229) .. (325.27,229) .. controls (327.64,229) and (329.55,230.91) .. (329.55,233.27) .. controls (329.55,235.64) and (327.64,237.55) .. (325.27,237.55) .. controls (322.91,237.55) and (321,235.64) .. (321,233.27) -- cycle ;

  \draw (103,36.5) node [anchor=north west][inner sep=0.75pt]   [align=left] {{\fontfamily{ptm}\selectfont Para-}};
  \draw (293,36.5) node [anchor=north west][inner sep=0.75pt]   [align=left] {{\fontfamily{ptm}\selectfont Ortho-}};
  \end{tikzpicture}
  \end{center}
\end{figure}
A grounded understanding of Heisenberg's image would have to wait the publication of his second paper, the one on the helium atom, \citet[]{RefWorks:1278}. We will come back to it in section \ref{1926b}.

In the case where spin is considered, a new interaction appears that breaks degeneracy. The spinless system had two spatial state functions (one symmetric for para-helium and one antisymmetric for ortho-helium). Now, we add four new spin-state functions (three symmetric and one antisymmetric) that must be multiplied with the spatial ones, giving a total of eight possible global state functions. Heisenberg represented this in a new image (Fig. \ref{HFig5}). Two possible global systems appear: one in which ortho-helium is a triplet and para-helium, a singlet (\textbullet), and another system in which ortho-helium is a singlet and para-helium a triplet (+). The reader must be cautious, though, because at this point, Heisenberg implicitly re-defines the meaning of the "systems" \textbullet \, and +. Up to that point (Fig. \ref{HFig4}) they had represented, respectively, para- and ortho-helium but now they come to represent the symmetry of the \emph{global state function} so that \textbullet \, represents the global antisymmetric state function and + represents the global symmetric state function. But, considering that the ground state was observed just in the system presenting singlets (para-helium), the system presenting triplets in the ground state, with global symmetric state function must be dismissed. The one that appears in nature was that represented by \textbullet \, so, with antisymmetric global state function and complying with Pauli principle. An attempt to clarify this quite puzzling arrangement is given in table \ref{space-spin}.
\begin{table}[htp]
  \centering
  \begin{threeparttable}
  \caption{Heisenberg's "triple-system" arrangement of $\bullet$/+ vs. Para/ortho vs. symm./antisymm.}
  \label{space-spin}
  \begin{tabular}{ccccccc}
  \hline\noalign{\smallskip}
  \multicolumn{2}{c}{Fig. \ref{HFig4}} & \multicolumn{1}{c}{|} & \multicolumn{4}{c}{Fig. \ref{HFig5}} \\
  \noalign{\smallskip}\hline\noalign{\smallskip}
      \textbullet/+
    & Spatial s.f.
    &
    & Spin s.f.
    & Global s.f.
    & \textbullet/+
    & Ortho-/para- \\
  \hline
      $+$
    & Antisymm.
    &
    & Symmetric
    & Antisymm.
    & $\bullet$
    & Ortho- \\
      $+$
    & Antisymm.
    &
    & Symmetric
    & Antisymm.
    & $\bullet$
    & Ortho- \\
      $+$
    & Antisymm.
    &
    & Symmetric
    & Antisymm.
    & $\bullet$
    & Ortho- \\
      $\bullet$
    & Symmetric
    &
    & Antisymm.
    & Antisymm.
    & $\bullet$
    & Para-\\
  \hline
      $\bullet$
    & Symmetric
    &
    & Symmetric
    & Symmetric
    & +
    & Para- \\
      $\bullet$
    & Symmetric
    &
    & Symmetric
    & Symmetric
    & +
    & Para- \\
      $\bullet$
    & Symmetric
    &
    & Symmetric
    & Symmetric
    & +
    & Para- \\
      $+$
    & Antisymm.
    &
    & Antisymm.
    & Symmetric
    & +
    & Ortho- \\
  \hline
  \end{tabular}
  \end{threeparttable}
\end{table}

Heisenberg now tries to argue that the reduction from 8 to 4 in the number of realizations of the global state functions implies a reduction of 2! in the "statistical weight", similar to the reduction appearing in Bose's way of counting. So, he argues, there is an actual relationship among quantum indeterminacy, Bose--Einstein statistics and Pauli's exclusion principle. The modern reader could be misled by the application of Bose--Einstein's statistics in the context of bound fermions. One can find several justifications (i.e. hypothetical Heisenberg's reasonings) that would justify his assertion. But, the fastest way to understand it is to recall Heisenberg's responses to Kuhn in 1963: "Somehow I always got mixed up between Bose's statistics which produced a different way of counting states, and the Pauli exclusion principle"  \citetext{AHQP, interview with Heisenberg, February 27, 1963}.\footnote{There is an interesting (and long) discussion about this topic in this particular interview.}

\subsection{The quantitative viewpoint}\label{1926b}

This section presents an independent assessment of Heisenberg's quantitative approach based on his paper but considering also the more elaborated versions of \comentario{Hans Bethe}\citet{RefWorks:1411} and \citet[]{RefWorks:1400}.\footnote{Readers interested in a historical account of Bethe's contributions to multi-electron atom studies will find valuable the biographic work by \comentario{Schweber's 2012}\citet{RefWorks:RefID:1209-schweber2012nuclear}.} The justification of the following assertions will drive the current section.
\begin{itemize}
  \item In his paper on the helium atom \citet[]{RefWorks:1278}, tried to ignore electronic orbits for the sake of calculations. He did not entirely dismiss them. Using Schrödinger solutions for two hydrogen-like electrons as a starting point for perturbation analysis allowed Heisenberg to consider the quantized kinetic energy just to the zeroth-order approximation, so that only potential energy was needed in the first approximation. Notice that the reliance of potential energy solely on intercharge distance simplified Heisenberg's argument for electron exchange, rendering considerations of momentum change superfluous. This facilitated him to continue speaking the language of orbits as an analogy\footnote{Sentences like "the position [...] relative to the 'orbit radius'" or "the two electrons [...] execute exactly the same motions (though in different phase) in the course of time" relay on the images of orbits in the classical sense.\label{0592}} despite his having previously discarded the "mechanical equations" of classical mechanics in his 1923 work on the helium atom \citetext{\citealt{RefWorks:1393}}. But, coherent with his claims, Heisenberg's "orbits" \emph{were not} of the kind one could expect from classical mechanics. Two main differences made it impossible to consider them as classical orbits:
  \begin{itemize}
    \item In the first place, he was using Schrödinger solutions for the hydrogen atom and those solutions did not represent a trajectory at all.
    \item In the second place, Heisenberg's electrons were presented as exchanging places without radiating. In our opinion, this "place exchange" was introduced by Heisenberg as a reminiscence of statistical mechanics where magnitudes were taken to be averages over ensembles of particles that had lost their identity (due to their assumed indistinguishability) in the sense that the average did not change when any two particles were considered to exchange places in the formulation.
  \end{itemize}
  \item Though indeterminacy has become a common argument against classical orbits, the radical novelty of Heisenberg's 1926 proposal, compared to his 1923 analysis, was not as much the introduction of indeterminacy as it was the statement of the symmetrization postulate and the idea of superposition, that emerged from Schrödinger wave mechanics, as a new form of particle indistinguishability.
\end{itemize}

Heisenberg tackles some calculations for systems with two electrons (He, Li\textsuperscript{+}, Be\textsuperscript{++} etc.) in \citetitle{RefWorks:1278}, \citet[]{RefWorks:1278}. His immediate goal is to apply first-order perturbation theory starting from a $H_0$ Hamiltonian that corresponds (nuclear motion neglected) to the simple ("unperturbed") hydrogen atom to eventually calculate an average (first-order) perturbation energy in the form $E_0 + E_1$ where $E_0$ (the energy of the zeroth-order approximation) is estimated from a solvable Hamiltonian and $E_1$ is determined to be
\begin{equation*}\label{0375}
E_1=\int U_0 H_1 \overline{U_0} \, d \tau
\end{equation*}
where $U_0$ is the zeroth-order wave function (i.e. the first term in the power series expansion $u=\sum_{n=0}^{\infty} \lambda^n U_n$ of a wave function $u$ assumed to solve a certain Hamiltonian), $\overline{U_0}$ its complex conjugate and $H_1$ the disturbing Hamiltonian.

The need to find a Hamiltonian symmetric in the potentials of the electrons in order to apply a systematic perturbation theory seems to have led Heisenberg to the following scheme. He starts by assuming two "hydrogen-like electrons" with interaction. He then, decomposes the corresponding energy into an "undisturbed potential" $V$ and a "disturbance potential" $W$
\begin{equation*}
-\frac{e^2 Z}{r_1}-\frac{e^2 Z}{r_2}+\frac{e^2}{r_{12}} = V + W
\end{equation*}
($r_i$ is the distance of the respective electrons to the nucleus and $r_{12}$ is the distance between them; $e$ is the electron charge). The potential $V$ for the whole atom is the sum, $V=V(r_1)+V(r_2)$, of the two electron potentials where
\begin{equation}\label{H1}
V(r)=-\frac{e^2Z}{r}+f(r), \; f(r)=\left\{\begin{array}{lll}
e^2 / r_0 & \text { for } & r<r_0, \\
e^2 / r & \text { for } & r>r_0
\end{array}\right\}. 
\end{equation}
The disturbance potential turns out to be
\begin{equation}\label{0382}
W=-\frac{e^2 Z}{r_1}-\frac{e^2 Z}{r_2}+\frac{e^2}{r_{12}}-V=\frac{e^2}{r_{12}}-f(r_1)-f(r_2). 
\end{equation}
Notice that Heisenberg deals with the excited atom which, due to the excited electron being more distant from the nucleus than the ground state one, allows him to choose the electron potential based on the existence of a threshold value $r_0$ for the radii. In the course of calculation this threshold $r_0$ cancels out due to its occurrence with different signs in \eqref{0403} below.

Heisenberg suggests thinking of his potential \eqref{H1} as caused by a uniformly charged spherical shell of charge $e$ and radius $r_0$. This can be misleading if taken literally. \comentario{Bethe, in 1933} \citet{RefWorks:1411} keeps the same formal expression for the potential energy but avoids the comparison and just mentions the shielding without reference to any geometry. Moreover, \comentario{Bethe's 1957 book}\citet[]{RefWorks:1400} do not even use Heisenberg's potential \eqref{H1} but simply accept the shielding and use a modified perturbation theory adapted to unsymmetrical potentials.

To allow for the use of Schrödinger wave functions, Heisenberg considers two independent hydrogen-like problems, one for an inner electron in the ground (1\textit{s}) state and other for an outer electron in some excited ($n$) state. Schrödinger's then recent work on the hydrogen atom led Heisenberg to the following set up. One defines, namely, for the inner electron, $V_0=-\frac{Ze^2}{r}+\frac{e^2}{r_0}$ and then considers the problem
\begin{equation*}
  \left(-\frac{1}{2} \Delta + V_0 + \overline{V(r)-V_0} \right)\psi = E_\text{inner}\psi,
\end{equation*}
where the bar over the second and third terms represents its average (Mittelung). This is essentially an approximation of the hydrogen problem due to the presence of the potential energy $-\frac{Ze^2}{r}$ in $V_0$, but the whole potential $V(r)$ is considered by including its average difference from $V_0$. The fact that the average is simply a number leads to (allowing for constants to depend on the units taken)
\begin{equation}\label{0403}
E_\text{inner}=-\frac{RhZ^2}{n^2}+\frac{e^2}{r_0}+\overline{f(r)-\frac{e^2}{r_0}},
\end{equation}
with $n=1$ and similarly for the outer electron
\begin{equation*}
E_\text{outer}=-\frac{Rh(Z-1)^2}{n^2}+\overline{f(r)-\frac{e^2}{r}}.
\end{equation*}
Giving a statistical meaning to the wave function,\footnote{The statistical interpretation of the wave function is commonly credited by historians to \citet[]{RefWorks:509}, in a paper published before Heisenberg's one. It can be worth noticing that Heisenberg's paper was received before Born's but published later and that Heisenberg applied the statistical interpretation to energy, not position as Born did. Discerning priority, however, is not our purpose here.} the averages are taken in the quantum mechanical sense, i.e., in the following expression for the zeroth-order approximation of the energy of the total atom, they would be written as they are in terms (a) and (b)\footnote{Heisenberg's original papers use two different notations for the state functions; in the first paper, \citet[]{RefWorks:1264}, discussed in the previous section, he uses $\varphi_n^a$, $\varphi_m^b$ where $n, m$ represent the states and $a,b$ represent the electrons; in the second paper, \citet{RefWorks:1278}, the electrons are identified by the radii of the wave function resulting from the solution of the (hydrogen-like) zeroth-order approximation problem, denoted as $\psi_n(r_1)$, $\psi_m(r_2)$, etc. We have preserved his notation to allow for a better confrontation with Heisenberg's corresponding papers, so the notation in this section differs from the previous one.}
\begin{equation*}
\begin{aligned}
E_0 &= Rh\left(- \frac{Z^2}{n_1^2}-\frac{(Z-1)^2}{n_2^2}\right)+\frac{e^2}{r_0} \\
    &+ \underbrace{\int |\psi_{n_1 l_1 m_1}(r_1)|^2\left(f(r_1)-\frac{e^2}{r_0}\right) \, d \Omega_1}_{\text{(a)}} + \underbrace{\int |\psi_{n_2 l_2 m_2}(r_2)|^2\left(f(r_2)-\frac{e^2}{r_2}\right) \, d \Omega_2}_{\text{(b)}}.
\end{aligned}
\end{equation*}
The Hamiltonian of the joint problem is the sum of the Hamiltonians of two hydrogen problems having the same eigenvalues. Then, the solutions include not only $$\psi_{n_1 l_1 m_1}(r_1)\psi_{n_2 l_2 m_2}(r_2)$$ but also the solution with permuted levels $$\psi_{n_2 l_2 m_2}(r_1)\psi_{n_1 l_1 m_1}(r_2).$$ This is coherent with Heisenberg's assumptions for the symmetry of the energy matrix stated in his previous paper (as discussed in section \ref{1926a}). This fact blurs the link between the electrons and their energy levels in a way similar to the finding that equation \eqref{0233} retains its value under permutation. The meaning of this is expressed by Heisenberg as electrons changing places, and can also be interpreted in terms of the indistinguishability of the electrons involved.\footnote{Unreasonable as it can sound that electrons can exchange places, it is worth noting that, months before Heisenberg's paper was published, a similar argument had been raised by N. R. Campbell who suggested, perhaps more as a philosopher than as a physicist,
\begin{quote}
    Particles moving with "uniform" velocity or oscillating in fixed orbits are, in our view, undergoing fortuitous transitions between the points of their paths, the intervals between these transitions being wholly irregular, but possessing a statistical average related simply to the distance covered in the transition. \citetext{\citealt[p. 1111]{RefWorks:RefID:507-campbell1926c.}}
\end{quote}}
Once these solutions have been suggested, any linear combination of them would also be a solution. To formally apply perturbation theory as indicated at the beginning of this section, one has to keep in mind that the final (global) eigenfunction needs to be antisymmetric, as seen in table \ref{space-spin}, so the spatial state function has to be symmetrized (i.e. a symmetric or an antisymmetric function). Following the procedure anticipated in his previous paper (see section \ref{1926a}), Heisenberg reduces the script notation so that $n_1 l_1 m_1 \equiv v$ and $n_2 l_2 m_2 \equiv w$ and writes, for the zeroth-order approximation eigenfunction, the possibilities
\begin{equation}\label{0424}
  U_{0\pm} = \frac{1}{\sqrt{2}}\left( \psi_v(r_1) \psi_w(r_2) \pm \psi_w(r_1) \psi_v(r_2) \right).
\end{equation}
Applying first order perturbation theory, one has
\begin{equation*}
  E_1 = \int\int U_{0\pm} W \overline{U_{0\pm}} \, d\Omega_1 \, d\Omega_2.
\end{equation*}
Two properties play a role: the orthonormality of the state functions and a permutation symmetry property of the product of the eigenfunctions $\psi_i$: that the following expressions labeled a and b (respectively a$'$ and b$'$) give the same results when used under the integral sign,\footnote{We will see in sect. \ref{discussion}-\nameref{discussion} that these conditions are equivalent to the condition that the integral at equation \eqref{0233} retains its value when $a$ and $b$ are exchanged.}
\begin{equation*}
    \underbrace{|\psi_v(r_1)\psi_w(r_2)|^2}_\text{a} \qquad \underbrace{|\psi_w(r_1)\psi_v(r_2)|^2 }_\text{b},
\end{equation*}
\begin{equation*}
    \underbrace{\psi_v(r_1) \psi_{w}(r_2) \overline{\psi_{w}(r_1)} \; \overline{\psi_v(r_2)}}_{\text{a}'} \qquad \underbrace{\psi_{w}(r_1) \psi_v(r_2) \overline{\psi_v(r_1)} \; \overline{\psi_{w}(r_2)}}_{\text{b}'}.
\end{equation*}
Developing the integrals that appear one has
\begin{equation}\label{0427}
  E = E_0 + E_1 = \underbrace{-Z^2 - \frac{(Z-1)^2}{n^2} + H^1(vw,vw)}_{\tilde{E}} \pm \underbrace{H^1(vw,wv)}_{K}.
\end{equation}
where
\begin{eqnarray} \label{0440}
  H^1(vw,vw) = \int \int W \psi_v(r_1) \; \psi_w(r_2) \cdot \overline{\psi_v(r_1)} \; \overline{\psi_w(r_2)} \, d \Omega_1 \, d \Omega_2, \\ \label{0441}
  H^1(vw,wv) = \int \int W \psi_v(r_1) \; \psi_w(r_2) \cdot \overline{\psi_w(r_1)} \; \overline{\psi_v(r_2)} \, d \Omega_1 \, d \Omega_2 .
\end{eqnarray}

Interpreting that the energy arising from the symmetric space eigenfunction in \eqref{0427} corresponds to para-helium, according to table \ref{space-spin}, and the antisymmetric one to ortho-helium then follows Heisenberg's assertion that "$H^1 (vw,wv)$ gives half the term distance between $O$[rtho-] and $P$[ara-] in the perturbed system".

If the integrals are calculated for He and Li\textsuperscript{+} one can derive the theoretical values for the Rydberg correction and compare them to the experimental ones. Heisenberg found an agreement that he considered acceptable given the restrictions of the model and the accuracy of the measurements (see \citet[tables 2 and 4]{RefWorks:1278}). The calculations were not exempt from criticism and, shortly after the paper was published it stimulated responses from Paul Dirac, Albrecht Unsöld, William Houston, J. A. Gaunt, Egil Hylleraas and others. A refined version of Heisenberg's numerical values can be seen in \comentario{Bethe and Salpeter}\citet[table 4 p. 136]{RefWorks:1400}.

Heisenberg considers then the electrons' spin and, after a similar discussion to that of the first paper (\citet{RefWorks:1264} discussed above in section \ref{1926a}), justifies again the energy differences, the singlet/triplet system of para- and ortho-helium and the fact that the ground state occurs only in para-helium.

There is an additional section in which Heisenberg argues, on the basis of his theory, to justify the fine structure of He and Li\textsuperscript{+} working on the spin--orbit coupling with a vector model. At the time, He and Li\textsuperscript{+} were considered to have a fine structure comprising a doublet but it was not well understood why they did not present a triplet as was the case of the alkaline earth series. Heisenberg seemingly proved in the mentioned section that the apparent doublet corresponded to a theoretical triplet and that its appearance as doublet was probably due to limited empirical accuracy. We reemphasize his use of a vector model, which elicited responses at \comentario{Gaunt}\citet{RefWorks:1326, RefWorks:1325}, although we cannot make a precise assessment of that section because we have not been able to disentangle a particular step of Heisenberg's reasoning (specifically the step between equations (28) and (29) of his paper).

\section{Synthesis of Heisenberg's papers}\label{Synthesis}

The main lines traced so far can be summarized as follows. Heisenberg had detailed knowledge of the problems arising when dealing with Bohr's version of the helium atom. He had made some calculations previously and knew the models that had already failed (mainly at determining the ionization energy). He was also skilled in perturbation theory, which he had used to attack this same problem in 1923. Then, in a couple of years, a series of developments brought the following theoretical devices to hand:
\begin{itemize}
  \item the Bose--Einstein and Fermi ways of counting,
  \item the Pauli exclusion principle,
  \item the Uhlenbeck--Goudsmit spinning electron,
  \item his matrix mechanics and,
  \item the Schrödinger wave functions.
\end{itemize}

Heisenberg devised the opportunity of testing the whole set of achievements in a single problem: the theoretical explanation of para- and ortho-helium. Neither Bose--Einstein nor Fermi distributions, which had also been laid out at the time, were essential to Heisenberg's goals.

Applying perturbation theory as a guiding principle, Heisenberg decided to take advantage of the Schrödinger approach which Schrödinger himself had proven to be equivalent to matrix mechanics. Matrix mechanics, for its part, allowed a simple representation of the discreteness of the energy eigenvalues (matrix $\mathbf{W}$) which was useful, jointly with the concept of indistinguishability of particles, as a preliminary approach to the symmetric and antisymmetric forms of the eigenfunctions to be used in the quantitative treatment of the problem. Heisenberg's claimed common origin of Bose--Einstein statistics and the Pauli exclusion principle, probably mistakenly based on a reduction factor of $2!$ in the number of state functions, was hardly tenable, although the flaw did not damage his other findings.

As to the calculation itself, he focused on an excited atom so that the perturbation scheme would converge quickly, and he thus avoided long calculations of second- and third- order approximations. This allowed him the definition of a symmetric potential that included, though, a rather arbitrary "uniform spherical shell" at a certain distance to make sense of the shielding of the excited electron. He considered two hydrogen-like electrons with a pair of energy values inherited from Bohr's theory joint with a set of proper eigenfunctions borrowed from Schrödinger's wave mechanics.

Heisenberg's symmetrization of eigenfunctions accommodated both Pauli exclusion principle and Uhlenbeck--Goudsmit's spin, resulting in a globally antisymmetric eigenfunction. Then, the explicit formulation of Schrödinger's solution for the hydrogen atom wave function was used to calculate the corresponding integrals and obtain interaction energies suitable for calculating the Rydberg correction and testing for agreement with experiment.

\section{Quantum statistics in Spring 1926}\label{qstats}

After the initial development of matrix and wave mechanics in 1925 and the spring of 1926, Heisenberg had an obvious interest in consolidating the value of his \citetext{\citealt{RefWorks:58, RefWorks:59, RefWorks:60}} matrix theory as the scaffolding for the new quantum mechanics, but Schrödinger's wave mechanics was already set up and the equivalence of both theories had been proven by \comentario{Schrödinger}\citet{RefWorks:56}. Additionally, in 1924, a new basis for the justification of Planck's radiation law had been suggested by \comentario{Bose}\citet{RefWorks:1067}, leading to a different way of counting radiation quanta. \comentario{Einstein}\citet{RefWorks:1058} and \citeyearpar{RefWorks:1060} readily extended the method to ideal gases. As we will see in this section, the new counting was soon interpreted to treat the quanta or molecules as statistically dependent, and later to regard the radiation quanta (and, thus, the particles of a gas) as indistinguishable.

Then, when Heisenberg wrote his first paper on the two-particle system in May/June 1926, quantum statistics had already begun to work. Furthermore, very recently (in May 1926), Fermi's paper \citetext{\citealt{RefWorks:RefID:9-fermi1926zur}} in which he included the exclusion principle in the statistical treatment of an ideal gas had been published. Heisenberg, then on a pleasure trip, had visited Fermi in Rome around April, presumably when Fermi had already submitted his paper to \emph{Zeitschrift für Physik}. They had a conversation on this topic:
\begin{quote}
  He told me that when one applied Pauli's exclusion principle one got something which is somewhat related to the Bose statistics, but which is definitely different from Bose statistics. It was a kind of complement. He told me that the relation between his statistics and Bose's was something like plus and minus. So that I knew from Fermi. The paper had not appeared at that time. It must have been April of '26 and that certainly must have influenced my paper on this resonance business... Do I not quote my conversation with Fermi? \citetext{AHQP, Interview with Heisenberg, February 27, 1963}
\end{quote}
Therefore, Heisenberg was aware of these developments, and there is no doubt that he knew the works of Bose and Einstein, as it is central to the last section of the \emph{Drei-männer Arbeit} \citetext{\citealt{RefWorks:60}}, although usually attributed to Jordan \citetext{\citealt[p. 341]{RefWorks:RefID:19-duncan2023constructing}}. Nevertheless, as we have already noticed and we will argue in the following sections, Heisenberg showed, at the time, a somewhat limited understanding of the relationship between the two statistics, specifically between Bose's work and Pauli's principle.

In July 1924, Bose had published the paper \comentario{'Plancks Gesetz und Lichtquantenhypothese'}\citetitle{RefWorks:1067} \citetext{\citealt{RefWorks:1067}}. The paper was published in \emph{Zeitschrift für Physik} on the recommendation of Einstein himself. Bose's paper had been rejected by the editors of \emph{Philosophical Magazine}, whereupon Bose sent it to Einstein and asked for his help in getting it published. It is widely accepted that, at that time, Bose was not fully aware of the significance of the reasoning presented in the article \citetext{\citealt{RefWorks:RefID:34-bergia1987discovered}} but Einstein, who had spent almost twenty years thinking about the statistical properties of light quanta, immediately recognized its relevance and promoted its publication.

Bose's crucial contributions were essentially twofold. The first was to deduce the following factor from a purely quantum approach:\footnote{$\rho_\nu$ density of action, $E$ the mean energy of Planck's oscillators.}
\begin{equation}
    \rho_\nu \, d \nu=\frac{8 \pi \nu^2}{c^3} \, d \nu \, E.
\end{equation}
This factor appeared in Planck's radiation law in 1900 and in 1924 still had a classical foundation based on the counting of the eigenmodes of a cavity. To obtain the new deduction, Bose divided the phase space into cells of volume $h^3$. Considering the two polarizations, the number of cells corresponding to the frequencies between $\nu$ and $\nu+d \nu$ would be:
\begin{equation}\label{0086}
    A^s(\nu) = \frac{8 \pi \nu^2}{c^3} V \, d \nu^s.
\end{equation}
The second contribution, no less relevant, was to calculate the number of microstates to later maximize them under the pertinent constraints. Given the number of cells \eqref{0086}, the probability of a state would be given by:
\begin{equation}\label{0090}
    \prod_s \frac{A^s(\nu)!}{p_0^{s}!p_1^{s}!\cdots}
\end{equation}
where $p_k^s$ indicates the number of cells with $k$ light quanta in level $s$. For a fixed energy, maximizing this expression led to Planck's law.

Einstein soon saw the significance of Bose's contribution and extended his treatment to the material ideal gas. This is not the place to detail his work, which he published in three installments \citetext{\citealt{RefWorks:1058, RefWorks:1060, RefWorks:RefID:38-einstein1925zur}}. What matters to us here is that, in the second installment, he stated that Ehrenfest and some other colleagues made him see the statistical peculiarity that he had introduced using formula \eqref{0090}. Indeed, being well versed in statistical physics, Ehrenfest had warned him of the loss of statistical independence of molecules as, in fact, he had already pointed in Planck's treatment \citetext{\citealt{RefWorks:RefID:40-ehrenfest1914simplyfied}}. At that time, Ehrenfest considered this statistical behavior as proof that Planck's energy elements were not particles. But the attribution of statistical dependence to the molecules of a gas was a significant change. Einstein's illustration is exemplary.

Einstein tried to clarify the situation using again Planck's famous combinatorial formula:
\begin{equation}\label{0096}
    \frac{\left(n_s+z_s-1\right)!}{n_{s}!\left(z_s-1\right)!},
\end{equation}
where $n_s$ is the number of particles and $z_s$ the number of cells of energy $\epsilon_s$. Were the particles statistically independent, he should have used:
\begin{equation}\label{0100}
    z_s^{n_s} .
\end{equation}
Furthermore, the product $\prod_s z_s^{n_s}$ would have to be multiplied by the factor
\begin{equation*}
    \frac{n!}{\prod n_{s}!}
\end{equation*}
because the permutations among all the particles must be included in the total sum. Next, Einstein calculates the entropies with the two expressions and argues that, in the second case, the classical one, it is necessary to manually eliminate the factor $n!$ to obtain an extensive entropy. However, he follows, when that is done the expression no longer fulfills Nernst's theorem while, in contrast, with the new expression \eqref{0096} entropy fulfills both requirements. Certainly, the discussion about that $n!$  had been going on since the time of Sackur and Tetrode, and Einstein himself had given a presentation on the matter at the Prussian Academy \citetext{\citealt{RefWorks:RefID:21-monaldi2009note}}. We will see that Heisenberg will wrongly comment on this point when he obtains the requirement of symmetrization in the framework of matrix mechanics.

This lack of statistical independence was what led Einstein to propose the existence of a certain interaction between molecules, and therefore to resort to the hypothesis of Louis De Broglie. \citet{RefWorks:1458} have argued that formulas \eqref{0096} and \eqref{0100} lend themselves to more than just Bose's original treatment: to a corpuscular interpretation. While Bose put the emphasis on the priority of cells and states over particles, Einstein highlighted the lack of statistical independence among particles. But mere statistics did not seem, to Einstein, sufficient reason to characterize the quantum ideal gas \citetext{\citealt{RefWorks:1377}}. He looked for a wave interaction that would explain this statistical dependence between the molecules of the gas. It is known that Schrödinger showed that, indeed, a rigorous (but slightly adapted to a material gas) wave approach also led to the new statistics, and that his foundational papers on wave mechanics can be considered a consequence of these reflections on Einstein's theory and De Broglie's hypothesis.

The other quantum statistics, that of Fermi--Dirac, was also born, let us say, inadvertently \citetext{\citealt{RefWorks:RefID:9-fermi1926zur}} and far from the nerve centers of physics, although Fermi had visited Göttingen and Leiden before finding the new statistics. What matters most to us here is that, precisely to avoid putting statistical arguments at the center, he refused to follow Einstein's footsteps, although there is no doubt that he was inspired by Einstein's contribution. Although Fermi had also made forays into the field of ideal gases, he seemed especially interested (as Heisenberg would later be) in helium. This is meaningful because Pauli's principle was born in the context of spectroscopy, to order and structure atomic layers based on spectral observations. Fermi did not take advantage of the insightful analysis that Einstein had introduced to clear up once and for all the approach by which Planck in 1900 had obtained the black body radiation law for the first time. He started from the formula:
\begin{equation}\label{0114}
    \frac{z_s!}{n_{s}!\left(z_s-n_s\right)!}
\end{equation}
(we continue using Einstein's notation), which would be the equivalent of \eqref{0096} but including the restriction of Pauli's principle. In this way, Fermi did not resort to microstates of the \eqref{0090} type, one of Bose's discoveries and which emphasized the \emph{occupation of states} instead of the \emph{distribution of particles}. Furthermore, in order to identify and define temperature, and instead of applying Boltzmann's principle as Bose and Einstein did, Fermi preferred to consider the limit in which the gas behaved ideally. He finally obtained the distribution today known as Fermi--Dirac:
\begin{equation}
    n_s=z_s \frac{1}{\alpha^{-1} e^{\frac{\epsilon_s}{\kappa T}}+1},
\end{equation}
where $\alpha$ is a Lagrange multiplier.

This was the state of affairs in May 1926 with regard to quantum ideal gas statistics. Both treatments, Einstein's and Fermi's, were completely framed in the old quantum theory and hence were, of course, based on the existence of particles.

\section{Indistinguishability "à la Heisenberg"}\label{indalaH}

Heisenberg's paper does not focus on interpretative issues, but neither can it be said to avoid them completely. His aim, as we have already mentioned, was twofold: to give an account of the helium spectrum and to try to relate the results of Bose, Einstein and Pauli to the new mechanics. Although Heisenberg did not properly undertake a statistical treatment, he discusses how the new mechanics can deal with a system of identical particles. What interests us here is to see how Heisenberg interpreted his solution of the two-body problem. In Monaldi's opinion, the works of both Heisenberg and Dirac "reveal a firm reliance on the corpuscular interpretation model, which continued to dominate the theoretical imagination of the two theoreticians, notwithstanding their overthrow of the classical representation of motion and their formalistic stances." \citetext{\citealt[p. 144]{RefWorks:984}}. In the following sections, we will argue that we quite agree with this assessment and we will try to clarify the features of the corpuscules that they were relying on.

In the present section, we divide our analysis into three parts: how Heisenberg interpreted the mathematical result (superposition of states with permuted subscripts), how the status of the concept of "particle" turned out, and, finally, the relationship of his discovery to the Pauli principle and the contributions of Bose and Einstein.

\subsection{Exchanging places}\label{explaces}

It is known (and we will return to it below) that, months later, Dirac formulated the symmetrization requirement for a state vector---with results similar to those of Heisenberg when applied to a two-electron system---openly starting from the premise that the permutation among particles yields physically indistinguishable states. Heisenberg also considers physically indistinguishable states although he is not as explicit as Dirac. For instance, when he illustrates the resonance phenomenon with oscillators formed by charged particles, he does not consider the transitions that would have distinguished the electrons involved. Also the picture he gives (exchange of places) of particle identity is evidence that Heisenberg makes use of it. Nevertheless, in the main result of his treatment, what we now call the symmetrization of wave functions (a superposition of stationary states), the subscripts he used to label wave functions could no longer be attached to single individual particles. We will turn to this in some detail in sections \ref{Indist} and \ref{0695}.

It is commonly believed that, at least until the appearance of the paper on the uncertainty principle, Heisenberg advocated not creating images of atomic processes. In the papers on the helium atom, he provides, albeit tentatively, a visualization of the perfect identity between particles: that they are constantly exchanging places. He warns, however, that under no circumstances this exchange can be understood as a motion in space--time:
\begin{quote}
  [\ldots] it seems to me that one of the most important aspects of quantum mechanics is that it is based on the corpuscular concept of matter; of course, this is not a description of the motions of corpuscles in our usual space--time concepts. This could hardly be expected; because even if the corpuscles turned out to be singularities of the metric structure of space, as is the wish of the continuum theories, this would probably not be a description in our usual space--time concepts---unless one counts a space whose dimensional determination deviates significantly from the Euclidean one as an "ordinary" space. \citetext{\citealt[pp. 412-413]{RefWorks:1264}}
\end{quote}
Indeed, Heisenberg insists on the corpuscular nature of matter, but with nuances and properties yet to be discovered and determined. This is how he proposes to visualize identity:
\begin{quote}
  If one wanted to get a clear picture of the motion of the electrons in the atom that corresponds somewhat to the quantum mechanical solution, one would have to imagine that the two electrons periodically swap places in a continuous manner, analogous to the energy oscillations in the oscillator example mentioned above, where the period of this beat is given by the distance of the orthohelium term from the corresponding parahelium term. \citetext{\citealt[p. 421]{RefWorks:1264}}
\end{quote}
Although this is the only interpretation he gives, he does not devote many lines to it, nor does he go into details. As we see it, with this type of exchange of place something similar happens to the transitions between states, whose indefiniteness was already in Bohr's atomic model of 1913: its dynamics, its kinematics, are never treated, because the states have become the protagonists of the descriptions, not the processes of change of state or the motion of electrons. Nowhere do we find how this exchange takes place. Note, however, that the paragraph quoted above begins with "\textit{If one wanted} to get a clear picture..." (our emphasis). Such a comment is also found in the second paper on helium, where he presents explicit calculations on helium and other elements:
\begin{quote}
  The following calculations are intended to give an overview of the motions of the electrons in the atom mentioned, \textit{as far as one can speak of motions in quantum mechanics}, and they are intended to derive the spectrum of He and Li\textsuperscript{+} qualitatively and to a large approximation quantitatively from the laws of quantum mechanics. \citetext{\citealt[p. 499, our emphasis]{RefWorks:1278}}
\end{quote}
And he adds:
\begin{quote}
  In quantum mechanics, such an initial system would not represent a reasonable approximation, since the two electrons periodically swap places, which is why sometimes one electron and sometimes the other is under the influence of the charge $Z$ or $Z-1$. \citetext{\citealt[p. 500]{RefWorks:1278}}
\end{quote}
We think that this absence of explanation of the swapping process must be understood precisely as evidence that Heisenberg did not think it was anything real. It was just a way of maintaining the idea of corpuscle and, at the same time, trying to deprive it of individuality. Nor in the subsequent evolution of Heisenberg's ideas, for example in the hands of Heitler and London, is such a process ever described \citetext{\citealt{RefWorks:22}}. \citet{RefWorks:1270} tried, in his textbook, to explain Heisenberg's work \citetext{\citealt{RefWorks:1043}}, as follows:
\begin{quote}
  [\ldots] the system $a$ in passing from the state $m$ to the state $n$ is emitting just the frequency to be absorbed by $b$ to cause it to pass from the state $n$ to the state $m$. This is \textit{resonance} as understood in quantum mechanics. Resonance can only occur if the systems are in \textit{different energy states}, one in the state $m$, the other in the state $n$, a consideration which does not enter into the classical theory of resonance. Resonance in quantum mechanics is essentially different from resonance in the classical theory; it is this difference, which cannot be bridged by correspondence principle methods, which accounts for the failure of the earlier attempts to solve the helium problem. \citetext{\citealt[p. 219, emphases in the original]{RefWorks:1270}}
\end{quote}
Whatever Birtwistle meant by "classical theory of resonance", the failed attempts to account for the helium spectrum had little to do with that concept---mostly because quantum resonance was far from well defined before Heisenberg's papers (and neither was it for some time after).

\citet[p. 31]{RefWorks:22} argues that this idea allowed Heisenberg to connect with the atomic transitions, central to matrix mechanics. The electronic transitions between states, although also without an associated definite process (quantum jumps), were more intuitive than the exchange between particles. Hence, the introduction of the concept of resonance. That it is a mechanism completely linked to permutations is clear from the fact that it only takes place if the electrons are not in the same state. Heisenberg interpreted superposition in this way, although he was not the first to use this mathematical concept, proper to waves. Schrödinger had already used it to speak of different energies, and interpreted it as simultaneous excitations of eigenmodes \citetext{\citealt{RefWorks:1123}}. This interpretation, however, cannot simply be transferred to the superposition presented by Heisenberg for the first time, since the terms of the superposition only differ in permutations of particles. As we have said, Born's probabilistic proposal was made in the same months (summer of 1926) and took a long time to be accepted \citetext{\citealt[p. 559]{RefWorks:RefID:19-duncan2023constructing}}.

\subsection{A new atomism}

This period coincides with the formation and interpretation of the concepts of the new mechanics. For example, and crucial for our discussion, the probabilistic interpretation of the wave function. That is, giving it a corpuscular interpretation and turning waves into probability amplitudes.

We have seen how there is no doubt that Heisenberg persisted in using corpuscular images of matter. Probably spurred on by the success of Schrödinger's proposal, he made it clear that, however fuzzy the idea of the corpuscle was becoming, it remained in the new mechanics. In fact, the enthusiasm for Schrödinger's waves can be said to have waned in the second half of 1926, to which Heisenberg himself certainly contributed with his criticisms. In July, Heisenberg and Schrödinger had a "public" discussion at a colloquium in Munich where they openly disagreed, and the first paper on the helium atom is the first publication in which Heisenberg criticized Schrödinger's proposal. He pointed out, among other problems, that Schrödinger waves were not defined in ordinary space.

Heisenberg initially did not take a dim view of Schrödinger's proposal, but gradually regretted that it was sold as a return to earlier conceptions. It was clear to him that, whether waves or particles, the new vision had to break with traditional images:
\begin{quote}
  Even if a consistent wave theory of matter in ordinary three-dimensional space could be developed, according to the program of de Broglie and Einstein, this would hardly provide an exhaustive description of the atomic processes in our ordinary space--time concepts. \citetext{\citealt[p. 412]{RefWorks:1264}}
\end{quote}

The philosophical position or, rather, its fluctuations in these aspects have been studied in detail \citetext{\citealt{RefWorks:38, RefWorks:31}}. According to Camilleri, Heisenberg did not deny the possibility of visualizing atomic processes based on principles, but was forced by circumstance. Moreover, in April 1926, he had a conversation with Einstein (who had invited him to Berlin to discuss the new theory) that plunged him into confusion \citetext{\citealt[p. 32]{RefWorks:31}}. Heisenberg believed that he had emulated the methodology that had led Einstein to redefine space and time with relativity. It was not so clear what might or might not be observable \emph{a priori}, and only a finished, complete theory could provide an answer. Heisenberg's proposed image apparently contradicted the positivism he established in 1925, but he did not give it much importance nor developed it further. Not without reason, Pauli accused him of being little "unphilosophical" because he paid "no attention to clear presentation of the basic assumptions and their relationship to previous theories".\footnote{Pauli to Bohr, 11 February 1924. In \citet[p. 25]{RefWorks:31}.}

In a talk he presented in Düsseldorf on 26 September at the annual meeting of German scientists, Heisenberg gave an initial assessment of the novelties introduced by the new mechanics \citetext{\citealt{RefWorks:45}}. Heisenberg affirmed that the degree of reality of atomic corpuscles was very different (and less) than that of everyday objects. He specified the aspects which called into question the classical idea of corpuscle, but described the corpuscular constitution of matter as evident. There were three "limitations" of the reality of corpuscles:
\begin{itemize}
  \item Attempts to explain microscopic phenomena had failed because it was not possible to visualize atoms nor atomic processes.
  \item Quantum mechanics defined quantities analogous to classical ones, such as position or momentum but, while energy could be associated with stationary states, for example, observable quantities associated with corpuscles were not directly visualizable either. One could not directly measure those magnitudes (such as energy or momentum) of a corpuscle.
  \item To account for a system of several bodies, it had to be assumed that they were indistinguishable, that they were constantly exchanging their positions. In the helium atom, for instance, it did not make sense to distinguish the interior electron from the exterior one.
\end{itemize}
Heisenberg concluded by appealing to De Broglie's original idea, to which he believed matrix mechanics was a worthy heir. Just as photons and electromagnetic waves can be compatible, so must electrons and wave functions be compatible. Now, the analogy between photons and electrons does not imply that matter has a wave nature but, rather, that both photons and electrons share a mixed, corpuscular/wave nature. It is clear that light quanta do not behave like ordinary objects because they have to be compatible with typical wave phenomena such as interference. In November he wrote to Pauli that, with the new mechanics "[\ldots] one no longer knows what the words \textquoteleft wave' and \textquoteleft particle' mean".\footnote{Heisenberg to Pauli, 23 November 1926. In \citet[p. 68]{RefWorks:31}.}

\subsection{Connection with the new statistics}

Although Heisenberg's paper is not strictly statistical, one of its declared goals is to establish a connection with previous work by Bose, Einstein and Pauli. It is not a statistical paper because temperature and thermal equilibrium do not appear, and neither does any thermodynamic quantity. However, one of Heisenberg's most relevant findings is how indistinguishability is to be treated in quantum theory, and this will have consequences for the counting of states that do, of course, affect statistical physics. Although, in his first paper, Heisenberg was wrong in his explanation of Bose's counting, his symmetrization assumption, also used by Dirac in a somewhat different way, allows the Bose--Einstein and Fermi counting to be deduced from the treatment of systems of identical particles in quantum mechanics. In this first paper, we can read:
\begin{quote}
  If only one (\textbullet) of the two systems occurs in nature, this gives rise, on the one hand, to a reduction of the statistical weights in the sense proposed by Bose; on the other hand, if the system is chosen correctly, Pauli's prohibition of equivalent orbits is automatically fulfilled. \citetext{\citealt[p. 422]{RefWorks:1264}}
\end{quote}
He later remarks: the statistical reduction corresponding to the Bose--Einstein counting is $n!$. As we argue at section \ref{discussion}--\nameref{discussion} and \comentario{Monaldi} \citet{RefWorks:984} also pointed out on different grounds, Heisenberg's alleged connection between Bose and Pauli seems to be a misinterpretation. His understanding of the statistical achievements of the old quantum theory appears to be limited at this stage. Just as Fermi did not \citetext{\citealt{RefWorks:1458}}, neither did Heisenberg go into the detailed explanations that Einstein himself had provided in his second paper on the quantum theory of the ideal gas.

In the previously mentioned Düsseldorf talk, Heisenberg still did not clarify the mess, although he explained the statistics of Bose and Einstein very clearly and was more cautious with the consequences of his treatment in that sense: "To what extent the Pauli prohibition, on the other hand, makes a modification of Einstein's statistics necessary could not be determined from this study" \citetext{\citealt[p. 993]{RefWorks:45}}. It seems beyond a doubt that he was not aware of Fermi's work.

That Heisenberg had carefully read Einstein's papers is shown by the comment he makes on the paradox formulated there and that is, as far as we know, scarcely commented \citetext{\citealt{RefWorks:1377}}. In a nutshell, it establishes that the new quantum ideal gas theory eliminates the Gibbs paradox but creates a new one. The classical paradox had to do with the extensivity of entropy, and was closely related to the dependence on the variable $N$. In the new statistics, this dependence disappears (in favor, we would say today, of the chemical potential), but in exchange, energy (and also pressure) becomes dependent on density. If we convert a mixture of two virtually identical gases into a single gas in a (mentally) continuous manner, there would be an abrupt change of energy (and thus pressure). \comentario{Ehrenfest and Uhlenbeck}\citet{RefWorks:ehrenfest1927einsteinschen} devoted an interesting paper to analyzing it. Describing his own standpoint, Heisenberg writes:
\begin{quote}	
  A paradox repeatedly emphasized by Einstein also has an analogue in our considerations: if the subsystems to be coupled are different from one another, classical statistics must apply to them; in principle down to infinitely small differences. Nevertheless, the counting for identical [sub]systems is completely different. For different [sub]systems, also according to the calculations carried out here, classical counting will always remain, since transitions between the $n$! subsystems occur; therefore, no subsystem can be excluded. However, the transitions become increasingly rare as the differences between the particles decrease. If the amplitudes corresponding to the transitions become smaller than a finite value defined by the sharpness of the state in question, there is a logical possibility of completely excluding the transitions and changing the counting. It should also be noted that, according to the considerations presented here, a finite interaction of the systems is a necessary prerequisite for changing the counting. If the periods of the energy impulses corresponding to the resonance effect are longer than the lifetime, the calculations presented above lose their meaning. \citetext{\citealt[p. 423]{RefWorks:1264}}
\end{quote}

Heisenberg's analogue of Einstein's emphasized paradox highlights the difficulties to proceed in a continuous manner from the classical treatment to a quantum mechanical one. In the quoted paragraph he equates, on the one hand, "classical counting" to "transitions" and to "place exchange with radiation" and, on the other hand, he equates "new way of counting" to "identical subsystems" and to "place exchange without radiation". The amplitudes (intensities) become smaller the more particles become similar, for then the place exchange with radiation turns into place exchange without radiation. His analogue of the paradox differs from Einstein's, which is intended to show that a mixture of two gases treated as quantum ideal gases leads to contradiction when one makes them equal.

It must be noted that in his second paper on resonance, \citetext{\citealt{RefWorks:1276}}, sent to the journal in December, Heisenberg showed knowledge of Dirac's \citeyearpar{RefWorks:RefID:605-dirac1926theory} paper and the connection of the symmetrization postulate with Bose--Einstein and Fermi statistics.

\section{Subsequent developments: Dirac}\label{Aftermath}
We now make a brief summary of several later papers connected with Heisenberg's ones and comment a bit more extensively on the one of \citet{RefWorks:RefID:605-dirac1926theory} already mentioned, as it is going to be relevant for our discussion on the topic of indistinguishability of particles.
\begin{itemize}
  \item \citetitle{RefWorks:RefID:1326-unsold1927beitrage} \citetext{\citealt{RefWorks:RefID:1326-unsold1927beitrage}}. Unsöld calculated, among other things, the ionization energy for the ground state of helium-like atoms using perturbation methods inspired by Heisenberg's works. The calculations came out too low, as Hylleraas would later recollect, because Unsöld used "independent spherical electronic charge distributions [that resulted] in an interelectronic energy unreasonably high" \citetext{\citealt[p. 425]{RefWorks:RefID:634-hylleraas1963reminiscences}}
  \item \citetitle{RefWorks:RefID:469-houston1927fine} \citetext{\citealt{RefWorks:RefID:469-houston1927fine}}. When Heisenberg wrote his papers on the helium atom, there were doubts about the reason why helium and Li\textsuperscript{+} differed in their spectral structure (doublets) from the alkaline earth series (triplets). Heisenberg concluded, from his calculations that, in the case of helium, the apparent doublet actually had to be a triplet. Houston's paper reported an experimental result that confirmed Heisenberg's theoretical argument of helium presenting a triplet instead of a doublet.
  \item \citetitle{RefWorks:1469} \citetext{\citet{RefWorks:1469}}. In this work, Hylleraas calculated the ground state energy of helium using a reference system given by the nucleus and the two electrons; the results enhanced those of \comentario{Unsöld}\citet{RefWorks:RefID:1326-unsold1927beitrage} at the abovementioned work.
  \item \citetitle{RefWorks:1326} \citetext{\citealt{RefWorks:1326}}. In this work, Gaunt called into question some of the assumptions that Heisenberg made when dealing with the helium atom. Among them was Heisenberg's use of a "vector model" for the triplets. He suggested that the calculation of the energies due to the spin were dealt with, in Heisenberg's works, by means of mechanical models. Gaunt, instead, made his calculations using the Dirac equation for the joint action of the electrostatic interaction and spin. The calculation, carried through by means of Schrödinger wave functions, was developed shortly after in \citet{RefWorks:1325}.
  \item \citetitle{RefWorks:1325} \citetext{\citealt{RefWorks:1325}}. In this paper, Gaunt gave the details of the calculations of the energy due to spin following the scheme presented in \citet{RefWorks:1326} (previously mentioned).
\end{itemize}
Heisenberg's paper on the helium atom was indeed well received by physicists. On 15th June 1926, a month before the paper was received at the journal, Kramers wrote the following to Schrödinger:
\begin{quote}
  Here in Copenhagen we are of course enjoying these days with Heisenberg's beautiful solution of the helium spectrum and his fundamental clarification of the branching theorem. These days I'm looking for a bit of mathematical methods to get through quantitatively a little faster. \citetext{\citealt[p. 270]{RefWorks:RefID:922-meyenn2011eine}}
\end{quote}

Heisenberg's resonance papers gave rise to the statement of the symmetrization postulate which led to a great simplification in the theory of elementary particles \citetext{\citealt[p. B249]{RefWorks:1456}}. The symmetrization postulate was developed on the idea of permutation symmetry understood as the fact that if a wave function, having an expression depending on physical quantities of the electrons, is a solution of a Schrödinger equation for the joint system, then the wave function with the parameters transposed will also be a solution.

This idea was also used in the above mentioned Dirac's \citeyearpar{RefWorks:RefID:605-dirac1926theory} work \citetitle[]{RefWorks:RefID:605-dirac1926theory} published on October 1926 and written independently of the papers by Fermi and Heisenberg. We highlight two aspects of his work that directly connect with our aim. First, Dirac's presentation of the symmetrization postulate \citetext{\citealt[\S 3]{RefWorks:RefID:605-dirac1926theory}} which was being published nearly at the same time in Heisenberg's paper. Second, Dirac's derivation of the Fermi theory for the ideal gas (made independently from Fermi) \citetext{\citealt[\S 4]{RefWorks:RefID:605-dirac1926theory}}.

In the case of Dirac, although the idea of permutation invariance could have had its origin in statistical mechanics, as in the case of Heisenberg, the \emph{need} of permutation invariance seems to have been an \emph{a priori} practical assumption. When dealing with similar particles he asks whether the state $(mn)$ of an atom and its permuted state $(nm)$ should count as two different states or as only one. After some argument, he stated:
\begin{quote}
  The two transitions are, however, physically indistinguishable, and only the sum of the intensities for the two together could be determined experimentally. Hence, \emph{in order to keep the essential characteristic of the theory} that it shall enable one to calculate only observable quantities, one must adopt the second alternative that $(mn)$ and $(nm)$ count as only one state. \citetext{\citealt[\S 3, p. 667. Our emphasis]{RefWorks:RefID:605-dirac1926theory}}
\end{quote}
This led him to the symmetrization postulate that wave functions must be symmetric or antisymmetric.

As to Dirac's theory for the ideal gas, he took as starting point the wave functions, in number $A_s$, for certain energy level $E_s$, being $N_s$ the number of molecules to be associated with the corresponding waves of that level $s$, then, calculating the probability as
\begin{equation}\label{0699}
W=\prod_s \frac{A_{s} !}{N_{s} !\left(A_s-N_s\right) !},
\end{equation}
and maximizing the entropy $S$
\begin{equation}\label{0703}
\begin{aligned}
S=k \log W=k \sum_s\left\{A_s\left(\log A_s-1\right)\right. & -N_s\left(\log N_s-1\right) \\
& \left.-\left(A_s-N_s\right)\left[\log \left(A_s-N_s\right)-1\right]\right\},
\end{aligned}
\end{equation}
he found the distribution for the $N_s$ to be given by\footnote{Here $\alpha$ is a Lagrange multiplier directly connected with the chemical potential.}
\begin{equation}
N_s=\frac{A_s}{e^{\alpha+E_s / k T}+1},
\end{equation}
which, unknown to Dirac, matched Fermi's distribution.\footnote{Dirac, although had knowledge of the existence of Fermi's work, was not conscious of his results \citetext{AHQP, Interview with P.A.M. Dirac, May 7, 1963}.\label{0877}} We point out here that Dirac assumed that the quantum principles then applicable to the atom were also applicable to the ideal gas. Distinct from Fermi, though, who based his arguments on the quantization of the energy of molecules behaving like harmonic oscillators, Dirac did not assume any particular way of quantization and developed his results directly from the number of wave functions expecting the molecules to be attached to them.

Let us say, on the one hand, that Fermi's \citeyearpar{RefWorks:RefID:10-fermi1926sulla} and \citeyearpar{RefWorks:RefID:9-fermi1926zur} articles on the new quantum statistics did not receive much international attention at the time of their publication. One of the main arguments justifying this low impact is that Fermi's formulation was based on old quantum theory. Since it still relied on Sommerfeld's quantization rules, it could be considered "outdated" by his contemporaries. Dirac himself recognized (see footnote \ref{0877}) that he had read Fermi's article but had completely forgotten about it because he did not find it related to his own work, which was already moving beyond old quantum theory. Coherently, in Dirac's \citeyearpar{RefWorks:RefID:605-dirac1926theory} work, published a few months after Fermi's \citeyearpar{RefWorks:RefID:9-fermi1926zur} article on quantum statistics came to light, there was no mention of Fermi's work, even though Dirac arrived at the same results as Fermi. In this controversy with the work developed by Dirac during the same period, it should be noted that Dirac was not actually trying to solve a specific problem when he encountered his quantum statistics (unlike Fermi, who was indeed trying to find a method for quantizing the ideal gas in accordance with the Nernst theorem). Dirac was instead attempting to find a more general formulation of the new quantum mechanics (in his case, the matrix formulation).

At the Como Conference in September 1927, although Fermi was not a speaker, his work echoed during the sessions and he made contributions and participated in the subsequent debates. In one them, Fermi explained the foundation of his statistics and the differences from Einstein's work, discussing the existence of two types of particles: those obeying Bose--Einstein statistics and those obeying his own statistics, pointing out that the relationships between these two statistics had already been clarified by Heisenberg, Dirac, and Wigner \citetext{\citealt[]{RefWorks:1415}}.

Heisenberg's ideas, on the other hand, were as fruitful as they were polysemic. \citet{RefWorks:22}, \citeyearpar{RefWorks:47} careful and brilliantly studied their evolution up to the notion of exchange forces, so important in quantum mechanics. Heitler and London were the first to apply symmetry to explain homopolar bonds, acknowledging their debt to Heisenberg's work and proposing an exchange of place between electrons. Heisenberg himself held quantum exchange interactions responsible for the magnetic order in his theory of ferromagnetism. Oppenheimer also used the same ideas, but added interpretative nuances to study collisions between atoms and electrons. Finally, Heisenberg again made use of exchange forces in the first explanation of nucleon cohesion. Heisenberg gave two interpretations of this proton--neutron interaction. On the one hand, the neutron and proton are constantly changing places; on the other hand, an electron is going back and forth between two protons (thus changing charge). The latter explanation gave rise to succeeding theories of forces in which a virtual particle is responsible for an interaction \citetext{\citealt{RefWorks:22,RefWorks:47,RefWorks:49,RefWorks:48}}.

\section{Discussion}\label{discussion}
Some elements for the discussion have been introduced in the previous sections. These elements are complemented here with some considerations as necessary to justify our claims as presented in section \ref{introduction}--\nameref{introduction}:\footnote{The comment at sect. \ref{ac}--\nameref{ac} applies to this section \ref{discussion}--\nameref{discussion}.}

\subsection{Indistinguishable/identical particles}\label{Indist}

Heisenberg's \citeyearpar{RefWorks:1264} first paper mentions that the component systems (the electrons) are identical. He argued on an analogy between the electrons and the classical coupled harmonic oscillator where identity is understood to mean that the oscillators have equal mass and the same fundamental frequency. He used this analogy in the first and second sections of the paper. Toward the end of the second section and before moving to the helium atom in the third, he introduced, by using the idea of particle indistinguishability, the symmetrized and antisymmetrized wave functions to conclude, together with his previously introduced matrix formulation, the non-combination of the para- and ortho-helium spectra.

For our purposes, and to avoid an ambiguous or excessively philosophical idea of indistinguishability, we will restrict our discussion to "observational indistinguishability". This could be defined in the following way.\footnote{For a discussion of the term with enriched terminology and an analysis of philosophical implications see \citet{RefWorks:RefID:24-dieks1990quantum}. Observationally indistinguishable particles are sometimes referred to as \emph{permutation-invariant} particles; throughout this work we have used this second denomination.}
 Particles $u$ with state vector $|u\rangle$ and $v$ with state vector $P|u\rangle$ (i.e. the state vectors differ only by a permutation $P$) are deemed \emph{observationally indistinguishable} if
\begin{equation}\label{0491}
\langle u|A| u\rangle=\left\langle u\left|P^{-1} A P\right| u\right\rangle.
\end{equation}
This, and other similar definitions, assume the wave function to be eventually inserted as a weight factor in the integral representing the measurement of a certain quantity which remains formally unaltered under permutation. We remark that the analysis of the permutation invariance \emph{of the state function} is not enough to determine the physical meaning of the permutation invariance that quantum mechanics demands. One has to apply the state function to calculate a specific observable and then check the calculated values against the measured ones.

One sees that, less formally, Heisenberg's equation \eqref{0233} has the same purpose as eq. \eqref{0491}, i.e. to state that the integral being zero implies that the exchange integral is real and this, in turn, implies that the particles are permutation-invariant. Only after the second paper was published, though, when applied to the Hamiltonian representing the inter-electronic interaction, did the equation attain its full significance.

Now that the concept is defined, it is worth pointing out that Heisenberg's assertions about electrons exchanging places can be better understood in terms of permutation invariance. It is similar to the case of classical statistical mechanics where a combinatorial permutation did not mean the particles physically exchanging all their dynamical quantities (position, momentum). An electron exchange in quantum mechanics, if understood in terms of permutation invariance, also does not mean a physical exchange but is a calculating device. A significant outcome of such a physical electron exchange, were it to happen, is that it would demand emission of radiation. As already mentioned at section \ref{explaces}--\nameref{explaces}, \citet[p. 219]{RefWorks:1270} incorporated Heisenberg's results and considered the problem to be solved by assuming that the emission of one particle gets compensated by the absorption of the other, much in the same way that one could understand resonance in a classical sense.

Heisenberg's formulation, though, did not involve the distributions of particles used in statistical mechanics because his statistical weights were not based on the number of eigenfunctions, as was the case of Dirac, but were instead based on the values of the eigenfunctions, so that the pretended factor $1/N!$ to justify indistinguishability was superfluous. As we argue below, his approach led to a very different kind of indistinguishability, more connected with what is now called superposition than with what was understood by the term in classical statistical mechanics or even in the then-new statistical mechanics of Bose--Einstein and Fermi.

\subsection{Superposition vs. indistinguishability and indeterminacy in the forsaking of orbital motions}\label{0695}

It is at times useful to present as one of the characteristics of quantum mechanics the fact that the uncertainty principle prevents a precise determination of the electron paths within an atom \citetext{\citealt[Ch. IX]{RefWorks:1457}}. It comes as no surprise, as Heisenberg's indeterminism principle is widely considered the fundamental physical lemma that epitomizes the spirit of the statistical outlook of quantum mechanics \citetext{\citealt[See][]{RefWorks:van1929statistical}}.

Given that Heisenberg, as mentioned above, had already suggested in his joint 1923 paper with Born that no reasonable orbital model was tenable, and given that his paper on the uncertainty principle, \citet[]{RefWorks:1217}, would not be published until a year later, one could be tempted to seek in these specific papers on resonance some argument \emph{against} orbits. This, however, is not the case.

As has been argued at section \ref{1926b}, Heisenberg's use of Schrödinger's solution for the Keplerian Hamiltonian as the starting point for perturbation theory facilitated his linking his language to orbital reasoning, even within the framework of quantum mechanics. Although the "orbits" were non-classical, he kept talking in terms of the classical concepts and even asserted explicitly that "the two electrons periodically change places" \citetext{\citealt[p. 220]{RefWorks:1278}} and that "the two electrons in the model [\ldots] execute exactly the same motions (although in different phase) in the course of time" \citetext{\citealt[p. 235]{RefWorks:1278}}. Both sentences convey the idea of orbiting electrons although their behavior faced difficulties similar to those of the Bohr and Schrödinger atoms.\footnote{It is worth noting that, in the course of time, perhaps in an attempt to avoid futile discussions, the language changed from "electrons change places" to "electrons change states" (can be seen in \citet[p. 132]{RefWorks:1400}). If one feels uncomfortable with the electrons exchanging places within an atom, the expression shift will not probably bring much relief, for the states were given by functions of the position coordinates of the electrons so the exchange of states without changing "places" could be found equally controversial.} Moreover, Heisenberg used explicitly a vector model to deal with the triplets of alkaline earth elements in the last section of his paper on helium.

But, given our claim that orbits were still present in Heisenberg's discussion of the helium atom,\footnote{See footnote \ref{0592}.} and given our current understanding that they are untenable, an explanation of the contribution of these specific papers to the dismissal of classical trajectories would be in order. A simple explanation, which can be seen as connected with the concept of indistinguishability, would be that no possible orbit was tenable that allowed the electrons a physical exchange of places without radiating. A careful reading of Heisenberg's text reveals that, aside from the use of Schrödinger's wave functions, this was the feature that most challenged classical orbits. The formal translation of this feature was the (now called) \emph{exchange integral} eq. \eqref{0441}. The integral had no obvious equivalent outside the quantum theory and its most usual interpretation today considers it as the frequency of exchange of the electrons' states. This requires the introduction of the time factors ($e^{-iEt/\hbar}$ for a given energy $E$) resulting from the separability of the time-dependent Schrödinger equation, so that the solutions at time $t=0$ would be, from eq. \eqref{0424},
\begin{equation*}
  \psi(0) = \psi_v\left(r_1\right) \psi_w\left(r_2\right) = \frac{1}{\sqrt{2}} \left( U_{0+} + U_{0-} \right)
\end{equation*}
and at time $t$, time factors included ($\tilde E$ and $K$ as in eq. \eqref{0427})
\begin{eqnarray*} \nonumber
  \psi(t) &=& \frac{1}{\sqrt{2}} \left( \psi_+(t) + \psi_-(t) \right) \\  \nonumber
          &=& \frac{1}{\sqrt{2}} \left( U_{0+} e^{-i \frac{(\tilde E + K)}{\hbar}t} + U_{0-} e^{-i \frac{(\tilde E - K)}{\hbar}t} \right) \\ \nonumber
          &=& e^{-i\frac{\tilde E}{\hbar}t} \left( \psi_v\left(r_1\right) \psi_w\left(r_2\right) \cos Kt - i \psi_w\left(r_1\right) \psi_v\left(r_2\right) \sin Kt \right),
\end{eqnarray*}
where the time-dependent state $\psi(t)$ is readily seen to be a superposition.

\subsection{Historical analysis}\label{Historicalsummary}

What has been said hitherto allows the following description of the historical development.

During the late 19th and early 20th centuries, Boltzmann's method of counting states was widely adopted. It was based on the assumption that, in a combinatorial sense, the component systems under study were distinguishable. The factor $1/N!$ which had been useful to obtain an extensive entropy (an entropy proportional to the number of molecules) was tolerated only if entropy differences where physically significant, but needed justification if entropy had to be considered an absolute quantity \citetext{\citealt{RefWorks:monaldi2022evolving}}. In 1924, when the paper by Bose and subsequent papers by Einstein appeared, the interpretation of the factor $1/N!$ was still an unsolved problem of statistical physics: if included, one obtained a desired extensive entropy but lost coherence with Nernst's theorem; if not included, one kept coherence with Nernst's theorem but lost the extensivity of entropy. Einstein's extension of Bose's ideas to gases made it apparent that this was not the case if, instead of Boltzmann's way of counting, one used that of Bose; in this later case, both requirements were fulfilled. That implicitly brought in the factor $1/N!$ which required dropping the statistical independence claimed for the molecules, i.e. dropping their distinguishability.

Heisenberg, conscious that indistinguishability was not only an accepted feature for a set of identical particles but a desired one, did not hesitate to use that feature. However, in contrast to Einstein in 1924, Heisenberg noticed the possibilities of the state functions of wave mechanics and he assumed that their values could be considered as statistical weights in averaging. He connected their fundamental indistinguishability with Pauli's exclusion principle by symmetrizing the Schrödinger solutions for the two-electron problem and, after considering also the spin state function, requiring the global state functions to be antisymmetric so that only para-helium, with an antisymmetric spin state function, was to be found in ground state. He also envisaged a connection between Bose--Einstein combinatorial indistinguishability and the Pauli exclusion principle which he could express by means of the state functions, arguing also on indistinguishability.

The Bose--Einstein way of counting had embedded in it the kind of combinatorial indistinguishability that differentiated it from Boltzmann's calculations for entropy. Einstein, by acknowledging in his second paper the statistical dependence between particles, had posited this indistinguishability as characteristic of the theory.

The situation was quite similar for the case of Fermi's way of counting. Fermi's way of counting, that happened to have embedded the same assumption on indistinguishability, had been published shortly before (in March of the same year of Heisenberg's paper, which was received in June). There, Fermi claimed that the kind of indistinguishability associated to the $N!$ factor could be dismissed when calculating entropy. His claim was correct, although from his point of view that specific characteristic was irrelevant for his calculation, as the process of maximization (which sought the point where the maximum was attained not the maximum itself) was not affected by the particles being distinguishable or not.

Heisenberg, at the time, had not yet gained a deep knowledge of Fermi's paper and had made just a first approach to Bose--Einstein's works. A better knowledge of both could have given him a hint that the combinatorial reduction of states implicit in both kinds of statistics had a deeper connection through a non-combinatorial form of indistinguishability rooted in the state functions of the system \emph{as a whole}. Heisenberg's assumption that indistinguishability should necessarily be reflected in a combinatorial reduction of $N!$ led him to link the antisymmetry of the state functions with the Bose--Einstein reduction probably because he was unaware that it could have led him also to Fermi reduction. Had he noticed that, then he perhaps would have gone a step further by correctly linking the symmetric state functions to the Bose--Einstein way of counting and the antisymmetric state functions to Fermi's way of counting.

This step was given by Dirac shortly after, shedding light on the relevance of the symmetry and antisymmetry of the state functions. Additionally, Dirac's focusing on the symmetrized and antisymmetrized state functions as the only acceptable ones made it evident that, although not yet termed that way, superposition was an unavoidable characteristic of quantum mechanics.

Heisenberg, though, did not at all use combinatorial statistics in his calculations. His aim in his work on helium was to justify that only para-helium was to be expected to be in ground state and, toward checking for experimental confirmation, to deal with first ionization considering a single excited electron while keeping the other electron in the ground state. He used the \emph{values} of the functions as statistical weights of the interaction energy instead of using their \emph{number} as a combinatorial device to calculate the thermodynamic weights, as Dirac did. In doing so, his statistical approach was nearer to the statistical interpretation of the wave function than it was to the statistical interpretation of entropy. Nevertheless, although not dealing with gases, he tried to establish an analogue of Gibbs paradox in which the difference between the gases was compared to the non-combining ortho- and para-helium \citetext{\citealt[p. 423]{RefWorks:1264}}.

\section{Conclusions}\label{Conclusions}

Heisenberg's papers on resonance supposed, indeed, a milestone of the new quantum mechanics. Among the relevant outcomes that the papers inspired were, to say some, the use of Pauli's exclusion principle in terms of the antisymmetry of the state functions, a justification of the "double spectra" of para- and ortho-helium, a functional relationship between the space and spin eigenfunctions in the context of Pauli's exclusion principle (i.e. the need of a global state function), a symmetrization requirement for the wave function, an approach to the calculation of the shielding of the electrons attending to its effect on the Rydberg correction term and a new application of perturbation theory techniques already used in astronomy (and, since Bohr, also in atomic physics) via the inclusion of Schrödinger's eigenfunctions as statistical factors. Despite Heisenberg's initial own claim, an accommodation of the Bose--Einstein way of counting within the frame of matrix mechanics was not firmly established.

Regardless of his rejection of mechanical models, Heisenberg kept using Keplerian rhetoric during the months following the publication of his celebrated paper on the \emph{Umdeutung}. His reasoning was not in any way comparable to that of classical orbital models in that he assumed that quantum jumps between physical locations (without radiation) were implicit even in stationary states. He thus partially retained orbital concepts despite the inapplicability of the "mechanical equations".

The symmetrization of the wave function---necessary to explain Pauli's principle---applied to the calculation of the energy implied a permutation invariance that was taken as an indication of a common frame for Bose--Einstein statistics and Pauli's principle. However, the energy was not calculated on the basis of the number of particles in a certain state but on the values of the state function itself, so the method had no evident connection with Bose's statistics. It was not that the electrons were indistinguishable in the combinatorial sense of Bose what suggested the notion of blurred orbits, but the fact that they were periodically exchanging places, an idea more connected with what was later known as superposition than it was with particle indistinguishability or with the idea of indeterminacy that Heisenberg would raise some months later.

\paragraph{\textrm{Acknowledgments}} We are greatly indebted to Joana Ibáñez. Our lively discussions and her many meaningful and demanding comments have been decisive for the final outcome. We thank Daniela Monaldi and Miguel Navascués for their insightful suggestions and comments. Maria Esteban's valuable contribution in double-checking some of our calculations has been very helpful. The basic ideas and the conclusions presented in this paper are our own. Consequently, any errors, misunderstandings, or incorrect assumptions that may be found within the work should be attributed solely to us. We acknowledge the support of the Spanish Ministry of Science and Innovation (PID2023-147710NB-I00).

\section{Declarations}\label{ac}
\paragraph{\textrm{Authors contributions}}
Although this work is a result of the joint effort of the whole team, due to the dynamics of our research some contributions of the second author deserve to be highlighted. A reference to this section will be made in such cases.

All authors were involved in the preparation of the manuscript and have read and approved the final version.

\paragraph{\textrm{Competing interests}}

The authors have no competing interests to declare that are relevant to the content of this article.

\nocite{RefWorks:RefID:37-kormos2015collected} \nocite{RefWorks:RefID:39-kormos2015collected} \nocite{RefWorks:RefID:1202-van1967sources}
\nocite{RefWorks:simon2006albert} \nocite{RefWorks:1376}

\section*{Archives}\label{archives}
\begin{tabular}{ll}
       \text{AHQP} & \text { Archive for History of Quantum Physics. American } \\
                   & \text { Philosophical Society, Philadelphia }
\end{tabular}

\printbibliography[] 
\end{document}
\typeout{get arXiv to do 4 passes: Label(s) may have changed. Rerun}